\documentclass[a4paper,12pt]{article}
\usepackage{jheppub} 
\usepackage{lineno}
\usepackage{physics} 
\usepackage{amsmath}
\usepackage{amsfonts}
\usepackage{amssymb}
 \usepackage{booktabs}
\usepackage{simpler-wick}
\usepackage{pdfpages}
\usepackage{mathtools}
\usepackage{lastpage}
\usepackage{dsfont}
\usepackage{tikz}
\usepackage{tikz-feynman} 
\usepackage{xcolor}
\usepackage{mdframed}
\usepackage{titlesec}
\usepackage{parskip}
\usepackage{enumitem}
\usepackage{microtype}
\usepackage{hyperref}

\usepackage{mathtools}
\usepackage{appendix}
\usepackage{comment}
\usepackage{tcolorbox}
\usepackage{tikz}
\usetikzlibrary{arrows.meta,decorations.pathmorphing,positioning}
\tcbuselibrary{skins,breakable,theorems}

\definecolor{PSblue}{RGB}{24,82,148}
\definecolor{PSgreen}{RGB}{39,128,76}
\definecolor{PSamber}{RGB}{180,120,0}
\definecolor{shadelight}{RGB}{242,246,255}
\definecolor{shadegreen}{RGB}{242,252,245}
\definecolor{shadeamber}{RGB}{255,252,238}

\newtcolorbox{PSresult}[1][]{
  enhanced, breakable, width=\textwidth,
  colback=shadegreen, colframe=PSgreen, arc=3pt,
  boxrule=0.8pt, left=8pt, right=8pt, top=6pt, bottom=6pt,
  fonttitle=\bfseries\color{black},
  title={Key result}, #1
}
\newtcolorbox{PSnote}[1][]{
  enhanced, breakable, width=\textwidth,
  colback=shadelight, colframe=PSblue!60, arc=3pt,
  boxrule=0.8pt, left=8pt, right=8pt, top=6pt, bottom=6pt,
  fonttitle=\bfseries\color{black},
  title={Remark}, #1
}
\newtcolorbox{PSphysics}[1][]{
  enhanced, breakable, width=\textwidth,
  colback=shadeamber, colframe=PSamber, arc=3pt,
  boxrule=0.8pt, left=8pt, right=8pt, top=6pt, bottom=6pt,
  fonttitle=\bfseries\color{black},
  title={Physical picture}, #1
}
\graphicspath{{figs/}}

\newcommand{\nc}{\newcommand}
\nc{\ba}{\begin{eqnarray}}
\nc{\ea}{\end{eqnarray}}
\nc{\bfk}{\bf{k} }
\nc{\bfq}{\bf{q} }
\nc{\rc}{\textcolor[rgb]{1.00,0.00,0.00}}
\nc{\bc}{\textcolor[rgb]{0.00,0.07,1.00}}

\renewcommand\thesection{\arabic{section}}

\title{RG-Flow Renormalized One-Loop Corrections to the Power Spectrum in USR Inflation}

\author{Haidar Sheikhahmadi$^{a}$}
\affiliation{$^{a}$School of Astronomy, Institute for Research in Fundamental Sciences (IPM), \\
P. O. Box 19395-5531, Tehran, Iran}
\emailAdd{h.sh.ahmadi@gmail.com; h.sheikhahmadi@ipm.ir}

\author{Amin Nassiri-Rad$^{b},^{a}$}
\affiliation{$^{b}$ Department of Physics, K.N. Toosi University of Technology,\\
 P.O. Box 15875-4416, Tehran, Iran}
\emailAdd{amin.nassiriraad@kntu.ac.ir}

\abstract{
The nature of one-loop corrections to long-wavelength CMB-scale modes in single-field inflation models with an intermediate ultra-slow-roll (USR) phase remains a subject of active debate. In this work, we perform a detailed investigation into the regularization and renormalization of these one-loop corrections to the curvature perturbation power spectrum. Employing a combined UV-IR regularization scheme within the renormalization group (RG) flow formalism, we compute the renormalized one-loop contributions, including those from the tadpole diagram, arising from both the cubic and quartic interaction Hamiltonians. This allows us to study meaningfully the running of the coupling constant aiming at removing the divergences appear in our study. We demonstrate that the fully regularized and renormalized fractional loop correction to the power spectrum is controlled by its peak value at the end of the USR phase, scaling as $\mathcal{P}_\mathrm{peak} \sim e^{6 \Delta N}$, where $\Delta N$ is the duration of the USR phase. This result confirms the original conclusion that loop corrections can become non-perturbatively large if the transition from the USR phase to the final slow-roll phase is instantaneous and sharp, potentially challenging the perturbative framework of such inflationary scenarios for primordial black hole formation.
}

\begin{document}
\maketitle
\tableofcontents

\section{Introduction}\label{intro00}

The calculation of one-loop corrections in single-field inflation models incorporating an intermediate ultra-slow-roll (USR) phase is currently a topic of significant debate~\cite{Kristiano:2022maq, Kristiano:2023scm, Firouzjahi:2023aum, Firouzjahi:2023bkt, Riotto:2023hoz, Riotto:2023gpm, Cheng:2021lif, Maity:2023qzw, Braglia:2024zsl, Choudhury:2023jlt, Choudhury:2023rks, Choudhury:2023hvf, Tasinato:2023ukp}. These models are particularly interesting because they provide a mechanism for generating primordial black holes (PBHs), which are potential candidates for dark matter~\cite{Ivanov:1994pa, Garcia-Bellido:2017mdw, Germani:2017bcs, Biagetti:2018pjj, Khlopov:2008qy, Ozsoy:2023ryl, Byrnes:2021jka, Escriva:2022duf, Hooshangi:2022lao, Firouzjahi:2023ahg, Kristiano:2024ngc, Raatikainen:2023bzk, Pi:2024jwt, Cheng:2023ikq, Inomata:2022yte}.

In the simplest realization, the inflationary dynamics consists of three consecutive phases: an initial slow-roll (SRI) phase, an intermediate USR phase, and a final slow-roll (SRII) phase. The USR phase is brief but engineered to enhance the curvature perturbation power spectrum by approximately seven orders of magnitude compared to its value on CMB scales. This enhancement is crucial for producing PBHs of the desired mass scales to serve as dark matter seeds~\cite{Kinney:2005vj, Martin:2012pe, Mohammadi:2018wfk}.

During a standard slow-roll phase, the curvature perturbation power spectrum is nearly scale-invariant with amplitude $\mathcal{P} \sim 2\times10^{-9}$. To form PBHs with masses relevant for dark matter, the power spectrum must be amplified by factors of $10^7$–$10^9$ on small scales. 

The debate was initiated by the claim in~\cite{Kristiano:2022maq} that short-wavelength modes exiting the horizon during the USR phase can induce significant one-loop corrections to long-wavelength CMB modes. The estimated fractional correction to the power spectrum scales as $\Delta \mathcal{P} / \mathcal{P}_\mathrm{CMB} \sim e^{6 \Delta N} \mathcal{P}_\mathrm{CMB}$, where $\Delta N \sim 2$–$3$ is the duration of the USR phase in e-folds and $\mathcal{P}_\mathrm{CMB} \sim 2 \times 10^{-9}$ is the amplitude of the power spectrum on CMB scales. It was argued that this correction could violate perturbative control, therefore it may challenge the consistency of the scenario for PBH formation in a perturbative scheme. This conclusion was supported and further elaborated using the effective field theory (EFT) of inflation in~\cite{Firouzjahi:2023aum, Firouzjahi:2023bkt}, where the necessary cubic and quartic Hamiltonians for a complete one-loop calculation were derived.

Subsequent works have presented contrasting views. Some studies argue that the dangerous loop corrections are suppressed by slow-roll parameters, especially if the transition to the final SR phase is mild, thus preserving perturbative control~\cite{Riotto:2023hoz, Riotto:2023gpm, Iacconi:2023ggt, Inomata:2024lud}. A third category claims that these corrections are volume-suppressed and therefore negligible~\cite{Fumagalli:2023hpa, Fumagalli:2024jzz, Tada:2023rgp, Kawaguchi:2024rsv, Inomata:2024dbr, Caravano:2024tlp, Caravano:2024moy}.

A central open question is whether proper regularization and renormalization procedures confirm the large loop corrections found in \cite{Kristiano:2022maq} or whether they suppress them. This work provides a solution by systematically applying a combined UV-IR cutoff scheme and renormalization within the in-in formalism.

In quantum field theory (QFT), the treatment of infrared (IR) and ultraviolet (UV) divergences is fundamental for obtaining physically meaningful results~\cite{DeWitt:1975ys, Birrell:1982ix, Fulling:1989nb, Parker:2009uva}. In this work, we address this issue directly by employing a cutoff regularization scheme to handle both IR and UV divergences systematically. Then to renormalize the divergences we use the renormalization group flow techniques \cite{Wallace:1974dx,Collecott:1974tg,Shafi:1976jk,Pisarski:1981hir,Peskin:1995ev}.
Now we turn our attention to renormalization group (RG)-flow approach. A key aspect of our renormalization procedure is the concept of the RG. After removing ultraviolet divergences by introducing counterterms, physical observables become independent of the cut-off $\Lambda$. This independence is encoded in the RG equations, which describe how the parameters of the theory (e.g., the Hubble scale $H$, the slow-roll parameters $\epsilon_i$, and the coupling constants, if exist) run with $M$, in which $M$ is the running parameter of the model.

In the context of USR inflation, loop corrections introduce logarithmic divergences of the form $\log(M/\Lambda)$ where $\Lambda$ is a UV cutoff (different from renormalization scale $M$). The RG flow allows us to absorb these logarithms into the running couplings, thereby resumming large logarithms that could otherwise spoil perturbative control. For example, the fractional correction obtained in Eq.~\eqref{fraction-cubic} contains a term $\log(-\Lambda\tau_i)$.

In a time-dependent background like inflation, the renormalization scale $M$ is often identified with the comoving momentum $k$ (or the Hubble scale $H$) \cite{Piazza:2009bp,Landim:2022jgr}. The RG flow then becomes a running with time or energy: as modes exit the horizon, the effective couplings evolve. In our calculation, we will use RG methods to:
\begin{itemize}
    \item Remove UV divergences systematically.
    \item Show that the physical (renormalized) one-loop correction is independent of the cutoff $\Lambda$.
    \item Relate the renormalized correction to the observed CMB power spectrum.
\end{itemize}

A key technical point is that the renormalization approach can also generate new operators not present in the original action. In our EFT of inflation, higher-order operators may receive loop-induced contributions, which may affect the consistency relations. We will not perform a full RG analysis here, but we will demonstrate that the dominant one-loop corrections are captured by the renormalization of the background parameters. A more thorough RG study, including the running of the sound speed and the slow-roll parameters, is left for future work.

\begin{figure}[ht]
    \centering
    \includegraphics[width=0.88\linewidth]{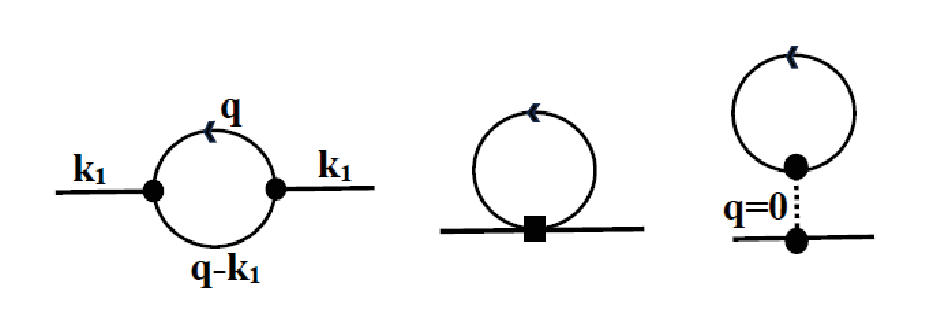}
    \caption{
        One-loop diagrams contributing to the power spectrum. These include contributions from cubic (order 3, black circles) and quartic (order 4, black squares) interaction vertices, as well as the tadpole diagram. In the cubic diagram, $\mathbf{k}_1$ denotes the external momentum, while $\mathbf{q}$ represents the internal momentum running in the loop. The quartic diagram has a similar structure. The tadpole diagram involves a zero-momentum ($q=0$) mode contraction.
    }
    \label{fig1}
\end{figure}

We utilize the in-in formalism~\cite{Weinberg:2005vy, Chen:2016nrs, Sheikhahmadi:2019xkx} to compute the loop corrections, which requires the interaction Hamiltonian up to fourth order. Our calculations, which incorporate a consistent UV-IR regularization scheme~\cite{Animali:2022lig, Ballesteros:2024zdp}, reveal divergences across the entire momentum integral range ($0$ to $\infty$).\footnote{Some previous studies evaluated integrals over restricted momentum ranges (e.g., from $k$ to $k_*$), which can significantly alter the resulting loop corrections by omitting contributions from certain modes.} We introduce explicit regulators: a lower cutoff $m_{IR} \to 0$ for IR divergences and an upper cutoff $\Lambda \to \infty$ for UV divergences. The results of these integrals are expanded as series in these cutoffs, allowing a systematic examination of their contributions.

As illustrated by the cubic diagram in Fig.~\ref{fig1}, the evaluation of momentum integrals for the internal momentum $\mathbf{q}$ requires careful treatment. Time integrals can also diverge and are handled using the $i\varepsilon$ prescription~\cite{Senatore:2009cf} and the Cauchy principal value (P.V.) method. The latter is particularly effective for managing divergences in the nested integrals that appear in the cubic loop corrections. UV divergences are subsequently removed by introducing appropriate counterterms.

We emphasise that our model is renormalizable, and appropriate counterterms will indeed be introduced to cancel the divergences of the type presented here (as well as other forms that may arise). The primary focus of this work is to identify the structure of the UV and IR infinities, implement a consistent regularization scheme, and then carry out a full renormalization procedure, including a renormalization group (RG) analysis of the running couplings. Similar approaches have recently been pursued in the literature: Ballesteros \emph{et al.}~\cite{Ballesteros:2024cef} and Braglia and Pinol~\cite{Braglia:2025cee} implemented such cancellations using counterterms, while Firouzjahi and Sheikhahmadi~\cite{Firouzjahi:2024faf,Firouzjahi:2023wbe} studied the structure of infinities and employed minimal subtraction to remove divergences.

Motivated by these considerations, we present a comprehensive re-analysis of loop corrections in the three-phase SRI-USR-SRII inflationary model. Our computation includes the complete set of one-loop diagrams: the quartic ($\mathbf{H}_4$) and cubic ($\mathbf{H}_3$) interactions, including the tadpole contributions. Both momentum and time integrals are subjected to rigorous regularization and renormalization to ensure consistency. By accounting for the full range of momenta and employing advanced regularization techniques, our study provides new insights into the behaviour of loop corrections in multi-phase inflationary models. We explicitly show that the regularized loop correction scales with the peak of the power spectrum at the end of USR, $\mathcal{P}_\mathrm{peak} \sim e^{6 \Delta N} \mathcal{P}_\mathrm{CMB}$, thereby confirming the conclusions of~\cite{Kristiano:2022maq, Firouzjahi:2023aum, Firouzjahi:2024sce}.

The remainder of the paper is organized as follows. Section~\ref{Mathematics} presents the theoretical setup, including the mode functions and the matching conditions. Section~\ref{sec:RG_review} provides a brief review on RG flow and its foundations. Section~\ref{sec:quartic} computes the quartic Hamiltonian contributions. Section~\ref{sec:cubic} deals with the cubic Hamiltonian, and Section~\ref{Tadpoles-Sec} covers the tadpole diagram. The appendices provide the explicit expressions for integrands and the zero-mode solutions. Finally, Section~\ref{conclusion0} summarizes our findings and discusses the implications for PBH formation.

\section{The Setup}
\label{Mathematics}

In this section we set up the theoretical framework for computing one-loop corrections to the curvature perturbation power spectrum in a three-phase USR inflation model. We begin by recalling the cubic and quartic interaction Hamiltonians derived from the effective field theory (EFT) of inflation \cite{Cheung:2007st, Cheung:2007sv}. In the decoupling limit, where gravitational backreaction is negligible, the Goldstone boson $\pi$ captures the dynamics of the adiabatic mode. The cubic and quartic Hamiltonians are given by \cite{Firouzjahi:2023aum}

\begin{align}\label{cubic-H3}
\mathbf{H}_3 &= - M_P^2 H^3 \eta \epsilon_H a^2 \int d^3 x \left[\pi \pi^{\prime 2} - \pi (\partial \pi)^2\right] \nonumber\\
&= - M_P^2 H^3 \eta \epsilon_H a^2 \int d^3 x \left[\pi \pi^{\prime 2} + \frac{1}{2} \pi^2 \partial^2 \pi\right], \\[4pt]
\mathbf{H}_4 &= \frac{M_P^2}{2 }\epsilon_H\int d^3x \Big[ \big( H^4 \eta^2 a^2 - \eta^{\prime} H^3 a \big)\pi^2 \pi^{\prime 2} 
+ \big( H^4 \eta^2 a^2 + \eta^{\prime} H^3 a \big)\pi^2 (\partial_i \pi )^2 \Big]. \label{quartic-H4}
\end{align}

Here $M_P$ denotes the reduced Planck mass, $a(\tau)$ the FLRW scale factor in conformal time, and $H$ the Hubble parameter during inflation (taken as approximately constant). The quantities $\epsilon_H$ and $\eta$ are the first and second slow‑roll parameters, respectively.

In the decoupling limit, the Goldstone mode $\pi$ is related to the curvature perturbation $\mathcal{R}$ by $\mathcal{R} = -H\pi + \mathcal{O}(\pi^2)$. This relation becomes exact on superhorizon scales once the system reaches the attractor phase. Thus, $\pi$ encodes the same physical information as $\mathcal{R}$, and the Hamiltonians above directly govern the interactions of curvature perturbations.

The slow‑roll parameters vary across the three phases of interest:
\begin{itemize}
    \item In the  SRI phase  (first slow‑roll), both $\epsilon_H$ and $\eta$ are small constants.
    \item In the  USR phase , we have $\epsilon_H = \epsilon_i (\tau/\tau_i)^6$ and $\eta = -6$, where $\tau_i$ denotes the conformal time at the start of USR. The duration of the USR phase in terms of e‑folds is $\Delta \mathcal{N} = \mathcal{N}(\tau_e) - \mathcal{N}(\tau_i)$, with $\tau_e$ the end of USR. From the scaling of $\epsilon_H$, one finds $\epsilon_e = \epsilon_i e^{-6 \Delta \mathcal{N}}$.
    \item In the  SRII phase  (final slow‑roll), the first slow‑roll parameter evolves as
    \begin{equation}\label{epsilonhminus}
        \epsilon(\tau) = \epsilon_e \left( \frac{h}{6} - \Big(1+\frac{h}{6}\Big)\Big(\frac{\tau}{\tau_e}\Big)^3 \right)^2,
    \end{equation}
    where $h \equiv -6 \sqrt{\epsilon_V/\epsilon_e}$, and $\epsilon_V$ is the asymptotic value of $\epsilon_H$ in SRII \cite{Cai:2018dkf}. 
    
    Here we present second slow-roll parameter $\eta$ as
    
    \begin{equation}\label{eta-2ndSR}
\eta(\tau)=-\frac{6(6+h)}{(6+h)-h\left(\frac{\tau_{\mathrm{e}}}{\tau}\right)^3}.
\end{equation}
    
    The parameter $h$ controls the sharpness of the transition to the final attractor. For a sharp transition one requires $|h|\gg 1$. Following \cite{Kristiano:2022maq}, we set $h = -6$, which yields $\epsilon_V = \epsilon_e$.
\end{itemize}
At transition point we have,
\begin{equation}
\begin{aligned}
&\eta=-6-h \theta\left(\tau-\tau_e\right) \quad \tau_e^{-}<\tau<\tau_e^{+} .\\
&\eta'=-h \delta\left(\tau-\tau_e\right), \quad \tau_e^{-}<\tau<\tau_e^{+}
\end{aligned}
\end{equation}

The term proportional to $\eta'$ in $\mathbf{H}_4$ produces a delta‑function contribution because $\eta$ jumps abruptly at the SRI-USR and USR–SRII transitions. This is crucial for capturing the correct matching of mode functions and will be carefully treated when evaluating loop diagrams.

\subsection{Mode functions}

To evaluate loop corrections using the in‑in formalism, we need the mode functions of $\mathcal{R}$ in all three phases. Starting from the Mukhanov–Sasaki equation for the curvature perturbation,
\begin{equation}\label{M-K}
(a^2 \epsilon \mathcal{R}')' + k^2 a^2 \epsilon \mathcal{R} = 0,
\end{equation}
and imposing Bunch–Davies initial conditions in the far past, the mode function in the first SR phase is
\begin{equation}\label{1stSRMode}
\mathcal{R}_k^{(1)} = \frac{H}{M_P\sqrt{4 \epsilon_i k^3}} (1 + i k \tau) e^{-i k \tau}.
\end{equation}

The transition from SRI to USR is assumed instantaneous at $\tau = \tau_i$. Imposing continuity of $\mathcal{R}$ and its derivative, the mode function in the USR phase becomes
\begin{equation}\label{USR-mode}
\mathcal{R}_k^{(2)} = \frac{H}{M_P \sqrt{4\epsilon_i k^3}} \left(\frac{\tau_i}{\tau}\right)^3
\Big[ \alpha_k^{(2)} (1+ik\tau) e^{-ik\tau} + \beta_k^{(2)} \, \text{c.c.} \Big],
\end{equation}
where “c.c.” denotes complex conjugation. The matching coefficients are
\begin{equation}\label{akbk-coef}
\alpha_k^{(2)} = 1 + \frac{3i}{2k^3\tau_i^3}\bigl(1+k^2\tau_i^2\bigr),\qquad
\beta_k^{(2)} = \frac{3i}{2k^3\tau_i^3}\bigl(-1+ik\tau_i\bigr)^2.
\end{equation}

Finally, matching at the end of USR ($\tau = \tau_e$), the mode function in the SRII phase is
\begin{equation}\label{SRIIMode}
\mathcal{R}_k^{(3)} = \frac{H}{M_P\sqrt{4 \epsilon(\tau) k^3}}
\Big[ \alpha_k^{(3)} (1 + i k \tau) e^{-i k \tau} + \beta_k^{(3)} \, \text{c.c.} \Big],
\end{equation}
where $\alpha_k^{(3)}$ and $\beta_k^{(3)}$ are given in Eqs.~\eqref{alpha-beta3} and \eqref{beta-beta3} in Appendix~\ref{AppA} 
\footnote{The integrations were performed using  \textit{Mathematica 14.3}.}. These coefficients are determined by the continuity conditions and depend on the sharpness parameter $h$.

\subsection{Regularization and the in‑in formalism}

Before computing loop diagrams, we summarize our technical setup:
\begin{enumerate}
    \item  UV–IR cutoff regularization: Divergences in the two‑point function are regulated by introducing a static IR cutoff $m_{IR} \to 0$ and a UV cutoff $\Lambda \to \infty$.
    \item  Time‑integral regularization and renormalization: Nested time integrals are regularized using the $i\varepsilon$ prescription and Cauchy priciple.
    \item  In‑in (Schwinger–Keldysh) formalism: All loop corrections are computed perturbatively using the Dyson series.
\end{enumerate}

The expectation value of an operator $\mathcal{O}$ up to second order in the presence of interaction Hamiltonian is
\begin{align}\label{DysonSeries}
\langle \mathcal{O}(\tau) \rangle_\Omega &\simeq \langle \mathcal{O}_I(\tau) \rangle_0 
 + 2 \,\mathfrak{Im}\! \int_{-\infty}^\tau d \tau' \, \big\langle \mathcal{O}_I(\tau) H_I (\tau') \big\rangle_0 \nonumber\\
 &\quad + \int_{\tau_0}^\tau d \tau_1 \int_{\tau_0}^\tau d \tau_2
 \langle H_I(\tau_1)\, \mathcal{O}_I(\tau)\, H_I(\tau_2)\rangle_0 \nonumber\\
 &\quad - 2 \,\mathfrak{Re}\! \int_{\tau_0}^\tau d \tau_1 \int_{\tau_0}^{\tau_1} d \tau_2
 \langle \mathcal{O}_I(\tau)\, H_I(\tau_1)\, H_I(\tau_2)\rangle_0,
\end{align}
where $\langle \cdots \rangle_\Omega$ and $\langle \cdots \rangle_0$ denote expectation values in the interacting and free vacua, respectively.

The first line of Eq.~\eqref{DysonSeries} corresponds to the free and quartic diagrams (insertions of $H_I$ with no nested integrals). The second and third lines yield the cubic diagrams (the final line is containing nested integrals). We will evaluate each contribution systematically.

With these tools, we are now ready to compute loop corrections. In the following sections, we first briefly review RG-flow techniques and then try to analyze the quartic diagram, then the more involved cubic contributions (which contain nested time integrals), and finally the tadpole diagram. The tadpole term requires solving for the zero mode of the Mukhanov–Sasaki equation; the explicit solutions, obtained by imposing CMB constraints and matching at the SRI–USR and USR–SRII transitions, are presented in Appendix~\ref{AppC}.

\section{Renormalization Group: a Brief Overview}
\label{sec:RG_review}
In this section we summarize the key ideas of the renormalization group  as developed in   \cite{Peskin:1995ev}. The RG provides a powerful framework for understanding how physical observables depend on the energy (or distance) scale at which they are measured, and it is essential for making sense of quantum loop corrections. Our discussion follows Chapters~10 to~13 of Peskin \& Schroeder (hereafter P\&S).

\subsection*{Renormalization conditions for massless theory}

For a massless scalar field, in flat spacetime, the on-shell conditions break down (they require a mass). P\&S instead impose conditions at the fixed spacelike reference momentum $p^2=-M^2$.
\begin{align}
  G^{(2)}(p)\big|_{p^2=-M^2} &= \frac{i}{p^2}, &
  G^{(4)}(p_i)\big|_{(p_i+p_j)^2=-M^2} &= -i\lambda,
  	\label{scaledef}
\end{align}
 The first equation fixes the field normalization, and the second defines the coupling $\lambda$. The parameter $M$ is arbitrary; shifting $M$ can be compensated by adjusting $\lambda$ and the field rescaling. We will come back to this important approach in next sections.

\subsection{Derivation of the Callan–Symanzik equation}

Write $\phi_0 = Z^{-1/2}\phi$ (bare vs. renormalized field) and $G^{(n)}_0 = Z^{n/2}G^{(n)}$. Bare Green’s functions depend on the bare coupling $\lambda_0$ and cutoff $\Lambda$, but  not  on $M$. Hence
\[
M\frac{d}{dM}G^{(n)}_0 = 0.
\]
Applying $M\,d/dM$ to $G^{(n)} = Z^{-n/2}G^{(n)}_0$ and keeping $\lambda_0,\Lambda$ fixed gives
\[
M\frac{\partial G^{(n)}}{\partial M}\bigg|_{\lambda_0,\Lambda}
= -\frac{n}{2}\frac{M}{Z}\frac{\partial Z}{\partial M}\,G^{(n)}
- M\frac{\partial\lambda}{\partial M}\bigg|_{\lambda_0,\Lambda}\frac{\partial G^{(n)}}{\partial\lambda}.
\]
Defining the RG functions
\begin{equation}
\beta(\lambda) \equiv M\frac{\partial\lambda}{\partial M}\bigg|_{\lambda_0,\Lambda},\qquad
\gamma(\lambda) \equiv \frac{M}{2Z}\frac{\partial Z}{\partial M},
\label{eq:beta_gamma_def_P&S}
\end{equation}
one arrives at the  Callan–Symanzik (CS) equation:
\begin{equation}
\left[ M\frac{\partial}{\partial M} + \beta(\lambda)\frac{\partial}{\partial\lambda} + n\gamma(\lambda) \right] G^{(n)}(\{x_i\}; M,\lambda) = 0.
\label{eq:CS_P&S}
\end{equation}

This equation states that a change in the renormalization scale $M$ can be compensated by a shift in $\lambda$ (via $\beta$) and a rescaling of each field (via $\gamma$), leaving all physical predictions unchanged.

\subsection{Computing $\beta$ and $\gamma$ at one loop}

P\&S show that the Callan–Symanzik functions are determined by the logarithmically divergent parts of the counterterms. For a theory with field-strength counterterm $\delta_Z$ and vertex counterterm $\delta_\lambda$, one has (to lowest order)
\begin{align}
\gamma(\lambda) &= \frac{1}{2}M\frac{\partial\delta_Z}{\partial M},\\
\beta(\lambda) &= M\frac{\partial}{\partial M}\!\left(-\delta_\lambda + \frac{1}{2}\lambda\,\delta_Z\right),
\end{align}

For massless $\phi^4$ theory in $d=4$, the one-loop propagator correction is momentum‑independent and does not contribute to $\delta_Z$; hence $\gamma = 0 + \mathcal{O}(\lambda^2)$. The four‑point counterterm gives
\[
\beta(\lambda) = \frac{3\lambda^2}{16\pi^2} + \mathcal{O}(\lambda^3).
\]

\subsubsection{Solution of the CS equation: the running coupling}

The CS equation can be solved by the method of characteristics. One defines the  running coupling  $\bar{\lambda}(p;\lambda)$ as the solution of the initial‑value problem
\begin{equation}
\frac{d\bar{\lambda}}{d\log(p/M)} = \beta(\bar{\lambda}),\qquad \bar{\lambda}(M;\lambda) = \lambda.
\label{eq:RG_equation_P&S}
\end{equation}
P\&S call this the renormalization group equation. An analogous formula holds for the four‑point function.

For $\phi^4$ theory: solving \eqref{eq:RG_equation_P&S} with $\beta(\lambda)=3\lambda^2/(16\pi^2)$ gives
\begin{equation}
\bar{\lambda}(p;\lambda) = \frac{\lambda}{1 - \dfrac{3\lambda}{16\pi^2}\log(p/M)}.
\label{eq:lambdabarP&S}
\end{equation}
It shows that the coupling grows with $p$ (IR free).

\subsection{The Nonlinear Sigma Model}
\label{sec:NLSM_review}

In this subsection we review the nonlinear sigma model (NLSM) in two spacetime dimensions. This model provides the simplest example of an asymptotically free quantum field theory in \(d=2\) and plays a crucial role in the understanding of critical phenomena via the \(d=2+\varepsilon\) expansion.

\subsection{Lagrangian and symmetry}

The NLSM describes an \(N\)-component unit vector field \(\mathbf{n}(x) = (n^1(x),\dots,n^N(x))\) constrained by \(\mathbf{n}^2 = 1\). The most general \(O(N)\)-symmetric Lagrangian with two derivatives is
\begin{equation}\label{eq:NLSM_Lag}
\mathcal{L} = \frac{1}{2g^2}\,(\partial_\mu n^i)^2,\qquad \sum_{i=1}^N (n^i)^2 = 1.
\end{equation}
The coupling \(g^2\) is dimensionless in \(d=2\). This Lagrangian can be interpreted as the low‑energy effective theory of \(N\) Goldstone bosons that arise when an \(O(N)\)-symmetric scalar field acquires a vacuum expectation value.

The NLSM describes the fluctuations of the direction of an order parameter whose magnitude is frozen. In the context of magnetism, \(\mathbf{n}(x)\) represents the local spin direction. The constraint \(|\mathbf{n}|=1\) forces the spins to lie on a sphere. The coupling \(g^2\) is inversely related to the stiffness: a small \(g^2\) corresponds to a large stiffness (ordered phase), while a large \(g^2\) corresponds to a floppy system (disordered phase).

\subsection{Parametrization and Feynman rules}

We solve the constraint by introducing \(N-1\) independent Goldstone boson fields \(\pi^k(x)\):
\begin{equation}
n^i = (\pi^1,\dots,\pi^{N-1},\sigma),\qquad \sigma = \sqrt{1-\boldsymbol{\pi}^2}.
\end{equation}
Expanding the Lagrangian in powers of \(\pi^k\) gives
\begin{equation}
\mathcal{L} = \frac{1}{2g^2}(\partial_\mu\pi^k)^2 + \frac{1}{2g^2}(\pi^k\partial_\mu\pi^k)^2 + \cdots,
\label{eq:NLSM_expanded}
\end{equation}
where the second term generates four‑point interactions and higher even powers. The Feynman rules are
\begin{itemize}
    \item Propagator: \(\langle \pi^k(p)\pi^\ell(-p)\rangle = \dfrac{i g^2\delta^{k\ell}}{p^2}\).
    \item Four‑\(\pi\) vertex (with two derivatives):
    \[
        \propto \frac{-i}{g^2}\bigl[(p_1+p_2)\cdot(p_3+p_4)\,\delta^{kl}\delta^{mn} + \text{two permutations}\bigr].
    \]
\end{itemize}

Because the Lagrangian is the most general \(O(N)\)-symmetric dimensionless Lagrangian, the theory is renormalizable in \(d=2\) with counterterms for \(g^2\) and the field strength.

\subsection{Computation of the Callan–Symanzik functions}

The renormalized Green’s functions satisfy the Callan–Symanzik equation
\begin{equation}
\left[M\frac{\partial}{\partial M} + \beta(g)\frac{\partial}{\partial g} + n\gamma(g)\right] G^{(n)} = 0.
\label{eq:CS_NLSM}
\end{equation}
P\&S compute \(\beta(g)\) and \(\gamma(g)\) at one loop by evaluating two simple Green’s functions.

\subsubsection{The one‑point function \(\langle\sigma\rangle\)}

Expanding \(\sigma = \sqrt{1-\boldsymbol{\pi}^2} = 1 - \frac12\boldsymbol{\pi}^2 + \cdots\), the one‑loop correction is
\begin{equation}\label{sigma-one}
\langle\sigma\rangle = 1 - \frac{N-1}{2}\,i g^2\int\frac{d^dk}{(2\pi)^d}\frac{1}{k^2-\mu^2} + \cdots,
\end{equation}
with \(\mu\) an infrared regulator. Evaluating the integral in \(d=2+\varepsilon\) dimensions, subtracting at the renormalization scale \(M\), and demanding that the result satisfy the CS equation yields 
\begin{equation}
\gamma(g) = \frac{(N-1)g^2}{4\pi} + \mathcal{O}(g^4).
\label{eq:gamma_NLSM}
\end{equation}

\subsubsection{The \(\pi^k\) two‑point function}

The one‑loop self‑energy from the four‑point vertex gives a logarithmic divergence. After renormalization, the propagator becomes
\begin{equation}
\langle\pi^k(p)\pi^\ell(-p)\rangle = \frac{i g^2\delta^{k\ell}}{p^2}\left[1 - \frac{g^2}{4\pi}\log\frac{M^2}{\mu^2} + \mathcal{O}(g^6)\right] + \cdots.
\label{eq:pi_prop_1loop}
\end{equation}
Applying the CS equation to this result and inserting \(\gamma(g)\) from Eq.~\eqref{eq:gamma_NLSM} gives the beta function:

\begin{equation}
\beta(g) = -\frac{N-2}{4\pi}\,g^3 + \mathcal{O}(g^5).
\label{eq:beta_NLSM_d2}
\end{equation}

For \(N=2\) the beta function vanishes at one loop. This is consistent with the fact that the \(O(2)\) NLSM can be rewritten as a free field \(\mathcal{L} = \frac{1}{2g^2}(\partial_\mu\theta)^2\) after the change of variables \(n^1 = \cos\theta,\; n^2 = \sin\theta\). A free theory has no coupling‑constant renormalization.

For \(N>2\), \(\beta(g)<0\): the theory is asymptotically free. The running coupling decreases at short distances and grows at large distances.

\section{Contributions from the Quartic Hamiltonian}
\label{sec:quartic}

In this section we compute the one‑loop correction to the power spectrum induced by the quartic interaction Hamiltonian $\mathbf{H}_4$ during the USR phase. The quartic diagram (see Fig.~\ref{fig1}) does not contain nested time integrals, so its evaluation is more straightforward than the cubic case. Nevertheless, careful regularization of both UV and IR divergences is required.

\subsection{General expression}

The one‑loop correction from $\mathbf{H}_4$ to the curvature perturbation two‑point function is given by
\begin{eqnarray}\label{USR-TPF}
  \langle {\mathcal{R}}_{k_1}^{(2)} (\tau_0)\,{\mathcal{R}}_{k_2}^{(2)}(\tau_0)\rangle_{\Omega_{H_4} }  
  \supseteq 2M_P^2 \, \delta_{k_1k_2}
   \int_{-\infty}^0 d\tau^\prime \int_0^\infty \frac{d^3{\mathbf{q}}}{(2\pi )^3}  
   \, \big( I_1 + 4I_2 + I_3 \big)(\tau^\prime)\,,
\end{eqnarray}
where $\delta_{k_1k_2} \equiv (2\pi )^3\delta^3(\mathbf{k}_1+\mathbf{k}_2)$. Here $\mathbf{k}$ denotes the long‑wavelength CMB mode (the external momentum), and $\mathbf{q}$ is the internal momentum running in the loop. The explicit forms of $I_1$, $I_2$, and $I_3$ are given in Eqs.~\eqref{I-one-three-1}–\eqref{I-one-three-3} in Appendix~\ref{AppA}. These integrands originate from the contractions of the quartic Hamiltonian with the mode functions.

\subsection{Contributions from the SRI and SRII phases}
\label{sec:SRI_SRII_contributions}

In this subsection we briefly discuss the expectation values
\[
\langle \mathcal{R}_{k_1}^{(2)}(\tau_0)\,\mathcal{R}_{k_2}^{(2)}(\tau_0) \rangle_{\Omega}
\]
for the slow‑roll phases, namely SRI (first slow‑roll) and SRII (final slow‑roll). 

For the SRII phase and with the specific choice of the transition parameter \(h = -6\) (the sharp transition limit adopted in this work), the expectation values vanish identically. This follows directly from Eq.~\eqref{eta-2ndSR} that the contribution is exactly zero.

For the SRI phase, after a careful evaluation of the one‑loop diagrams we obtain the following results. The contribution from the bulk of quartic Hamiltonian \(H_4\) is

\begin{equation}\label{SRI-H4}
\begin{aligned}
\langle \mathcal{R}_{k_1}^{(2)}(\tau_0)\,\mathcal{R}_{k_2}^{(2)}(\tau_0) \rangle_{\Omega_{\text{SRI}}} \big|_{H_4}
= \frac{i (\tau_{ini} - \tau_i)}{768M_P^4 k_{CMB}^3\, t_e^3 \epsilon_i^2} \Big[ & 3 \eta_{\text{SRI}}^2 H^4 \Lambda^2 \tau_e^3\tau_i \\
& - 2 (6 + \eta_{\text{SRI}}) H^4 \Lambda^4 \tau_i^4 (\tau_e^3 - 2 \tau_i^3) \Big],
\end{aligned}
\end{equation}
where $\tau_{ini}$ refers the initiation of inflation. Similarly, the contribution from the cubic Hamiltonian \(H_3\) reads

\begin{equation}\label{SRI-H3}
\begin{aligned}
\langle \mathcal{R}_{k_1}^{(2)}(\tau_0)\,\mathcal{R}_{k_2}^{(2)}(\tau_0) \rangle_{\Omega_{\text{SRI}}} \big|_{H_3}
= \frac{H^4 \tau_i^6\eta_{_{SRI}} \Big(1+348 \log\left(2-243 \log(3)\right)\Big)}{2560 M_P^4 k_{CMB}^3 \tau_e^6 \epsilon_i^2}+\mathcal{O}(\epsilon^2).
\end{aligned}
\end{equation}
One notices that to get the contribution of $H_3$ is a little tricky. According to Eq.~\eqref{DysonSeries} we must break the nested integrals to different intervals as follows

\begin{eqnarray}\label{Integralintervals}\nonumber
\int_{-\infty}^{\tau}d\tau_1\int_{-\infty}^{\tau_1} f(\tau_1,\tau_2)d \tau_2=\int_{-\infty}^{\tau_i}d\tau_1\int_{-\infty}^{\tau_1} f(\tau_1,\tau_2)_{_{SR-SR}}d \tau_2\\
+\int_{\tau_i}^{\tau_e}d\tau_1\int_{-\infty}^{\tau_i} f(\tau_1,\tau_2)_{_{SR-USR}}d \tau_2+\int_{\tau_e}^{0}d\tau_1\int_{-\infty}^{\tau_1} f(\tau_1,\tau_2)_{_{USR-USR}}d \tau_2,
\end{eqnarray}
where subscripts indicate the related mode functions for different eras of inflationary evolution.

The expression Eq.~\eqref{SRI-H4} is bare a result; It depends explicitly on the ultraviolet cutoff \(\Lambda\) (through powers \(\Lambda^2\) and \(\Lambda^4\)). This indicates that a naive hard cutoff regularization is not a reliable tool for handling momentum integrals in the SRI phase. As we will see, these divergences must be removed by a consistent renormalization procedure. After renormalization, the physical (cutoff‑independent) contributions from the SRI phase are suppressed and can be neglected compared to the USR results.

Then according to Eq.~\eqref{SRI-H3}, this result is the so called $\epsilon$ suppresed in inflationary contex and has no contribution to the final results. Therefore, confidently the contribution of SRI and SRII are negligible.
 
\subsection{Bulk contributions (USR phase)}

We first consider the bulk contributions coming from the USR phase, i.e., from the interval $\tau_i < \tau' < \tau_e$. Applying cutoff regularization (UV cutoff $\Lambda$, IR cutoff $m_{IR}$) and expanding around $\Lambda\to\infty$, $m_{IR}\to 0$, expanding in CMB modes, we obtain:

\begin{PSresult}[title={Bulk contributions from quartic Hamiltonian}, width=\dimexpr\textwidth+1.6cm\relax]
\begin{eqnarray}
\label{5-9Ha}
  \langle {\mathcal{R}}_{k_1}^{(2)} (\tau_0)\,{\mathcal{R}}_{k_2}^{(2)}(\tau_0)\rangle_{\Omega_{I1\text{bu}}}
  &=& - \frac{36H^4 e^{6 \Delta \mathcal{N}} \delta_{k_1k_2}}{32 M_P^4 \epsilon_{i}^2 k_{\mathrm{CMB}}^3}
  \Big[ 3\log (- \Lambda\tau_i) + 3\gamma_{\mathrm{EM}} - 2 + 3\log(2)\Big],\\
\label{5-10H}
  \langle {\mathcal{R}}_{k_1}^{(2)} (\tau_0)\,{\mathcal{R}}_{k_2}^{(2)}(\tau_0)\rangle_{\Omega_{I2\text{bu}}}
  &=& \frac{24H^4e^{6 \Delta \mathcal{N}}\, \delta_{k_1k_2}}{32 M_P^4 \epsilon_{i}^2 k_{\mathrm{CMB}}^3}
  \Big[ 3\log (- \Lambda\tau_i) + 3\gamma_{\mathrm{EM}} - 2 + 3\log(2)\Big], \\
\label{5-11H}
  \langle {\mathcal{R}}_{k_1}^{(2)} (\tau_0)\,{\mathcal{R}}_{k_2}^{(2)}(\tau_0)\rangle_{\Omega_{I3\text{bu}}}
   &=& - \frac{189H^4e^{4 \Delta \mathcal{N}}\, \delta_{k_1k_2}}{64 M_P^4 \epsilon_{i}^2 k_{\mathrm{CMB}}^3 }
  \Big[ 2\log (- \Lambda\tau_i) + 2\gamma_{\mathrm{EM}} - 3 + 2\log(2)\Big].
\end{eqnarray}
\end{PSresult}

Here $\gamma_{\mathrm{EM}} \approx 0.5772$ is the Euler‑Mascheroni constant, and $k_{\mathrm{CMB}}$ denotes the CMB pivot scale. The appearance of logarithmic terms $\log(-\Lambda\tau_i)$ is a characteristic feature of quantum loop corrections in de Sitter space. The dominant contributions scale as $e^{6\Delta\mathcal{N}}$, while the gradient term (the one proportional to $I_3$) scales only as $e^{4\Delta\mathcal{N}}$ and is therefore subleading for $\Delta\mathcal{N} \gtrsim 1$.

The factor $e^{6\Delta\mathcal{N}}$ arises from the growth of the mode functions during the USR phase. Since the power spectrum at the end of USR is $\mathcal{P}_{\mathrm{peak}} \sim e^{6\Delta\mathcal{N}} \mathcal{P}_{\mathrm{CMB}}$, the loop correction inherits this enormous enhancement. This is the origin of the potential breakdown of perturbation theory.
We must emphasis here that, we only keep the leading terms in whole calculations.

\subsection{Renormalization scale and interpretation of $\Lambda$}

It is important to clarify the role of the cutoff $\Lambda$, which is comoving, in Eqs.~\eqref{5-9Ha}–\eqref{5-11H}. The parameter $\Lambda$ is  not  a physical cutoff signalling unresolved divergences. Rather, in the spirit of effective field theory, $\Lambda$ should be understood as a  renormalization parameter. Logarithmic terms of the form $\log(-\Lambda\tau_i)$ are cured using the RG-flow mechanism. Then, the power law UV divergences will be absorbed into the counter terms.

The integrands in Eq.~\eqref{USR-TPF} are divergent. One must be cautious: Fubini’s theorem does not permit freely interchanging the order of momentum and time integrations, nor the order of summation and integration in divergent series. To avoid inconsistencies, we adopt a  cutoff regularization scheme  where we impose finite integration limits $\int_{m_{IR}}^\Lambda$ with $m_{IR}\to0$ and $\Lambda\to\infty$, both cutoffs are comoving. After evaluating the momentum integral, we expand the result in two complementary regimes:
\begin{enumerate}
    \item  UV limit  – expand around $\Lambda\to\infty$ to capture high‑energy behaviour.
    \item  IR limit  – expand around $m_{IR}\to0$ to capture long‑wavelength contributions.
\end{enumerate}
This systematic procedure allows us to disentangle UV and IR divergences and properly identify the renormalization scale $M$.

From an order‑of‑magnitude perspective, the result obtained for the USR phase is consistent with previous findings in the literature \cite{Kristiano:2022maq, Firouzjahi:2023aum, Firouzjahi:2024sce}.

\subsection{Transition contributions from $\eta'$}

In addition to the bulk contributions, we must account for the  localized source  at the end of the USR phase. This arises from the term proportional to $\eta'$ in $\mathbf{H}_4$ (see Eq.~\eqref{quartic-H4}), which produces a delta‑function $\delta(\tau-\tau_e)$ due to the abrupt change in $\eta$ at the USR–SRII transition. Evaluating these transition contributions yields:

\begin{PSresult}[title={Transition contributions from $\eta'$},width=\dimexpr\textwidth+1.6cm\relax]
\begin{eqnarray}
\label{5-9trans-H}
  \langle {\mathcal{R}}_{k_1}^{(2)} (\tau_0)\,{\mathcal{R}}_{k_2}^{(2)}(\tau_0)\rangle_{\Omega_{I1\text{tr}}}
  = \frac{12H^4 e^{6 \Delta \mathcal{N}}\, \delta_{k_1k_2}}{32 M_P^4 \epsilon_{i}^2 k_{\mathrm{CMB}}^3}
  \Big[ 3\log (- \Lambda\tau_i) + 3\gamma_{\mathrm{EM}} - 2 + 3\log(2)\Big],\\
\label{5-10trans-H}
  \langle {\mathcal{R}}_{k_1}^{(2)} (\tau_0)\,{\mathcal{R}}_{k_2}^{(2)}(\tau_0)\rangle_{\Omega_{I2\text{tr}}}
  = -\frac{24H^4e^{6 \Delta \mathcal{N}}\, \delta_{k_1k_2}}{32 M_P^4 \epsilon_{i}^2 k_{\mathrm{CMB}}^3}
  \Big[ 3\log (- \Lambda\tau_i) + 3\gamma_{\mathrm{EM}} - 2 + 3\log(2)\Big],\\
\label{5-11trans-H}
  \langle {\mathcal{R}}_{k_1}^{(2)} (\tau_0)\,{\mathcal{R}}_{k_2}^{(2)}(\tau_0)\rangle_{\Omega_{I3\text{tr}}}
  = - \frac{9H^4e^{4\Delta \mathcal{N}}\, \delta_{k_1k_2}}{32 M_P^4 \epsilon_{i}^2 k_{\mathrm{CMB}}^3}
  \Big[ 2\log (-\Lambda\tau_i) + 2\gamma_{\mathrm{EM}} - 3 + 2\log(2)\Big].
\end{eqnarray}
\end{PSresult}

Notice that the transition contributions have the same logarithmic structure as the bulk terms, but with different numerical coefficients. When summed together, the leading $e^{6\Delta\mathcal{N}}$ terms combine to produce the net correction.

\subsection{Final fractional correction to the power spectrum}

After combining the bulk and transition contributions for the leading terms (the $e^{6\Delta\mathcal{N}}$ pieces) and applying, the fractional correction to the power spectrum takes the form:

\begin{PSresult}[title={Final fractional correction from quartic Hamiltonian}]
\begin{eqnarray}\label{fraction-cubic}
\frac{\Delta \mathcal{P}}{\mathcal{P}_\mathrm{CMB}} 
\simeq -12\,\frac{\mathcal{P}_\mathrm{CMB}}{k_{\mathrm{CMB}}^3}\,e^{6 \Delta \mathcal{N}}  \Big[ 3\log (- \Lambda\tau_i) + 3\gamma_{\mathrm{EM}} - 2 + 3\log(2)\Big]\, ,
\end{eqnarray}
where $\mathcal{P}_\mathrm{CMB} = H^2/(4 M_P^2 \epsilon_i)$ is the tree‑level CMB power spectrum amplitude.
\end{PSresult}

As emphasised in \cite{Kristiano:2022maq}, for the specific choice $h = -6$ the contributions from the SRII phase are equal to zero. In this work we therefore restrict ourself to $h = -6$ case. A more general analysis with arbitrary $h$ will be presented in a future work, where the one‑loop contributions from the SRII phase will also be addressed.

\section{Contributions from the Cubic Hamiltonian}
\label{sec:cubic}

The cubic Hamiltonian $\mathbf{H}_3$ in Eq.~\eqref{cubic-H3} gives rise to one-loop corrections that are more involved than those from the quartic Hamiltonian, because they contain  nested time integrals  (see the Dyson series expansion, Eq.~\eqref{DysonSeries}). In this section we compute these contributions systematically, regularizing the divergences using the Cauchy principal value prescription.

\subsection{General expression from the Dyson series}

For the two-point function of curvature perturbations during the USR phase, the cubic contribution reads
\begin{eqnarray}\label{Two-Point-Cubic00}
\langle{\mathcal{R}}_{k_1}^{(2)} (\tau _0)\,{\mathcal{R}}_{k_2}^{(2)}(\tau _0)\rangle _{\Omega _{{H_3}}} &\equiv & 
\langle {\mathcal{R}}^{(2)}_{k_1k_2}(\tau _0)\rangle _{\Omega _{{H_3}}}
= \nonumber\\
&-& \int_{ -\infty }^{{\tau _0}} d {\tau _1}\int_{ -\infty }^{{\tau _1}} {d{\tau _2}}
\Bigg[ \int_0^\infty \frac{d^3{\bf{q}}}{(2\pi)^3} \Big(
\langle {\bf{H}}_3(\tau _1)\,{\bf{H}}_3(\tau _2)\,{\mathcal{R}}^{(2)}_{k_1k_2}(\tau _0)\rangle_0 \nonumber\\
&+& \langle {\mathcal{R}}^{(2)}_{k_1k_2}(\tau _0)\,{\bf{H}}_3(\tau _1)\,{\bf{H}}_3(\tau _2)\rangle_0 
\Big)\Bigg] \nonumber\\\nonumber
&+& \int_{ -\infty }^{{\tau _0}} d {\tau _1}\int_{ -\infty }^{{\tau _0}} d {\tau _2} 
\Bigg[ \int_0^\infty \frac{d^3{\bf{q}}}{(2\pi)^3} 
\langle {\bf{H}}_3(\tau _1)\,{\mathcal{R}}^{(2)}_{k_1k_2}(\tau _0)\,{\bf{H}}_3(\tau _2)\rangle_0
\Bigg] \, .\\
\end{eqnarray}
The nested time integrals in the first term (the “time‑ordered” piece) and the last term (the “anti‑time‑ordered” piece) are highly non‑trivial. In the USR phase, divergences arise at the upper  limits of integration.
In brief we can use the following approximations to carry out the integrals

\begin{enumerate}[leftmargin=2em, itemsep=2pt]
  \item \textbf{Large‑scale limit} ($k \ll k_{\rm peak}$): simplifies the mode
        functions and produces the scale‑invariant behavior discussed above.
  \item \textbf{Sharp‑transition limit}: decouples the USR phase from the SR
        phases and allows closed‑form evaluation of the time integrals.
  \item \textbf{Leading‑logarithmic approximation}: retains only the dominant
        UV‑divergent logarithms that feed into the RG running.
  \item \textbf{Cauchy principal‑value prescription}: resolves the ambiguity in
        the nested time integrals arising from the in‑in contour, as in the standard
        treatment of literature.
\end{enumerate}

\subsection{regularization via Cauchy principal value}

To regulate these infinities, we employ the well‑established  Cauchy principal value  prescription. This method introduces a parameter that controls the approach of the integration boundaries, consistently removing unphysical divergences while retaining the finite, physical contributions to the correlators. After regularization, the remaining finite part is independent of the regulator.

Substituting the cubic Hamiltonian $\mathbf{H}_3$ from Eq.~\eqref{cubic-H3} into Eq.~\eqref{Two-Point-Cubic00} and focusing on the USR phase ($\tau_i \leq \tau \leq \tau_e$), the one‑loop correction sourced by cubic interactions reduces to
\begin{eqnarray}\label{Two-Point-Cubic01}
\langle {\mathcal{R}}_{k_1}^{(2)} (\tau _0)\,{\mathcal{R}}_{k_2}^{(2)}(\tau _0)\rangle_{\Omega_{H_3}} 
= -8 M_P^4 \int_{\tau_i}^{\tau_e} d\tau_1 
\int_{\tau_i}^{\tau_1} d\tau_2 
\int \frac{d^3 \mathbf{q}}{(2 \pi)^3}\, 
\mathcal{G}\!\left(\tau_1, \tau_2 ; q\right)\, ,
\end{eqnarray}
where $\tau_i$ and $\tau_e$ denote, respectively, the beginning and end of the USR stage, and $\mathbf{q}$ is the loop momentum. The function $\mathcal{G}(\tau_1,\tau_2;q)$, defined explicitly in Eqs.~\eqref{G-cal1}–\eqref{G-cal3} in Appendix~\ref{AppB}, encapsulates the structure of the cubic interaction vertices as well as the propagators of the curvature perturbation. The integration domain is the triangular region $\tau_i \leq \tau_2 \leq \tau_1 \leq \tau_e$, reflecting the causal ordering of the Dyson expansion.

The function $\mathcal{G}$ contains three types of terms, corresponding to:
\begin{itemize}
    \item  Time‑time (t‑t) : both vertices involve $\pi \pi'^2$ (time derivatives).
    \item  Gradient‑gradient (gr‑gr) : both vertices involve $\pi (\partial \pi)^2$ (spatial derivatives).
    \item  Time‑gradient (t‑gr) : mixed contributions.
\end{itemize}
Each type must be evaluated separately. The momentum integral over $\mathbf{q}$ is performed first, followed by the nested time integrals. The Cauchy principal value prescription is applied to handle divergences at the integration limits.

\subsection{Regularized results}

After regularizing the infinities and assuming that divergences can be removed by suitable counterterms (renormalization), the one‑loop correction from the  time‑time (t‑t)  bulk cubic term is obtained as
\begin{eqnarray}\label{Two-Point-Result}
\langle {\mathcal{R}}_{k_1}^{(2)} (\tau _0)\,{\mathcal{R}}_{k_2}^{(2)}(\tau _0)\rangle_{\Omega_{H_3{t-t}}} 
= \frac{24 H^4\,e^{6\Delta \mathcal{N}}}{32 M_P^4  \epsilon_i^2\,k_{\mathrm{CMB}}^3}\,(1-6\Delta \mathcal{N})\,\delta_{k_1k_2}\, ,
\end{eqnarray}
where $\Delta \mathcal{N}$ is the number of e‑folds during the USR phase.

For the  gradient‑gradient (gr‑gr)  and  time‑gradient (t‑gr)  contributions, the results take a similar form but with different numerical coefficients:
\begin{eqnarray}\label{Two-Point-Grad-Result}
\langle {\mathcal{R}}_{k_1}^{(2)} (\tau _0)\,{\mathcal{R}}_{k_2}^{(2)}(\tau _0)\rangle_{\Omega_{H_3{gr-gr}}} 
= \frac{15 H^4\,e^{6 \Delta \mathcal{N}}}{32 M_P^4  \epsilon_i^2\,k_{\mathrm{CMB}}^3}\,\delta_{k_1k_2}\, ,
\end{eqnarray}
and
\begin{eqnarray}\label{Two-Point-Grad2-Result}
\langle {\mathcal{R}}_{k_1}^{(2)} (\tau _0)\,{\mathcal{R}}_{k_2}^{(2)}(\tau _0)\rangle_{\Omega_{H_3{t-gr}}} 
= \frac{3 H^4\,e^{6 \Delta \mathcal{N}}}{32 M_P^4  \epsilon_i^2\,k_{\mathrm{CMB}}^3}\,(35-24\Delta \mathcal{N})\,\delta_{k_1k_2}\, .
\end{eqnarray}

\begin{itemize}
    \item All three types scale as $e^{6\Delta\mathcal{N}}$, the same exponential factor that controls the peak power spectrum. This confirms that loop corrections are exponentially sensitive to the duration of the USR phase.
    \item The time‑time contribution contains a term linear in $\Delta\mathcal{N}$ (from the factor $1-6\Delta\mathcal{N}$), which can dominate for $\Delta\mathcal{N} \gtrsim 1$.
    \item The gradient‑gradient contribution does not have  linear growth in $\Delta\mathcal{N}$, while the time‑gradient term also contains a linear factor $(-24\Delta\mathcal{N})$.
\end{itemize}

\subsection{Final fractional correction from cubic Hamiltonian}

Collecting all cubic contributions (Eqs.~\eqref{Two-Point-Result}–\eqref{Two-Point-Grad2-Result}) and applying renormalization (which subtracts the UV divergences and absorbs them into the definition of cosmological parameters), the leading fractional correction to the power spectrum takes the form
\begin{PSresult}[title={Final fractional correction from cubic Hamiltonian}]
\begin{eqnarray}\label{fraction-cubic}
\frac{\Delta \mathcal{P}}{\mathcal{P}_\mathrm{CMB}} 
\simeq -108\, \frac{\mathcal{P}_\mathrm{CMB}}{k_{\mathrm{CMB}}^3}\,e^{6 \Delta \mathcal{N}} \Delta \mathcal{N}\, ,
\end{eqnarray}
\end{PSresult}
where we have kept the leading term proportional to $\Delta\mathcal{N}$ (the time‑time and time‑gradient contributions both produce such linear growth). The numerical coefficient $27$ arises from the combination of the three contributions.

This result matches the structure advocated in Refs.~\cite{Kristiano:2022maq, Firouzjahi:2023aum, Firouzjahi:2023bkt}, confirming the robustness of the regularized one‑loop enhancement in non‑attractor scenarios. For $\Delta\mathcal{N} \sim 2$–$3$, $e^{6\Delta\mathcal{N}} \sim 10^5$–$10^8$, the product $\mathcal{P}_\mathrm{CMB} e^{6\Delta\mathcal{N}} \Delta\mathcal{N}$ can be as large as $\sim 10^{-4}$–$10^{-1}$, potentially violating perturbative control unless further cancellations occur.

\section{Contributions from the Tadpole}
\label{Tadpoles-Sec}

We now turn to the contribution of the  tadpole diagram  (see Fig.~\ref{fig1}), which corresponds to the zero‑mode contribution in the interaction picture. Unlike the cubic and quartic diagrams, the tadpole involves contractions of a single interaction vertex with the zero‑mode of the curvature perturbation.

\subsection{Zero‑mode solutions and matching conditions}

To evaluate the tadpole terms, we make use of the Mukhanov–Sasaki (M‑S) equation for the zero‑mode solution $\mathcal{Q}_{k=0}$. By imposing the observational constraints from the CMB superhorizon modes at both the initiation and the end of inflation, together with the matching conditions across the SRI $\rightarrow$ USR and USR $\rightarrow$ SRII transitions, we obtain explicit solutions for the tadpole zero modes, as given in Eqs.~\eqref{ZeromodeSRI}–\eqref{ZeromodeSRII} of Appendix~\ref{AppC}.

The tadpole diagram represents a two‑point function with zero modes, see Fig.~\ref{fig1}, the third diagram. In the presence of a non‑attractor phase like USR, the zero mode can acquire a non‑trivial time dependence, leading to a non‑zero tadpole. This contributes to the two‑point function through contractions with the external legs.

\subsection{Contractions and the kernel $\mathcal{S}$}

Applying Wick’s theorem to the zero modes and considering all possible contractions arising from the Dyson expansion in Eq.~\eqref{DysonSeries}, we arrive at the following structure for one of the possible contractions:

\begin{eqnarray}
\label{eq:cubic-4th-combin}
& &\int_{ -\infty }^{{\tau _0}} d {\tau _1} \int_{ -\infty }^{{\tau _1}} {d{\tau _2}}\int_0^\infty \frac{d^3{\bf{q}}}{(2\pi)^3} \big(\langle \mathcal{O}_I(\tau) H_{I\mathcal{C}}(\tau_1) H_{I\mathcal{C}}(\tau_2) \rangle_\Omega \big) \nonumber\\\nonumber
&=& \int_{ -\infty }^{{\tau _0}} d {\tau _1}\int_{ -\infty }^{{\tau _1}} {d{\tau _2}}\int_0^\infty \frac{d^3{\bf{q}}}{(2\pi)^3} 
\wick{ \c1{\mathcal{R}_{k_{\mathrm{CMB}}}(\tau)} \c2{\mathcal{R}_{k_{\mathrm{CMB}}}(\tau)} 
\c1{\mathcal{R}_{q}(\tau_1)} \mathcal{R}_{q}^{\prime}\c2{(\tau_1)} 
\c4{\mathcal{R}_{q}^{\prime}}(\tau_1) \c4{\mathcal{R}_{k}}(\tau_2) 
\c3{\mathcal{R}_{k}^{\prime}}(\tau_2) \c3{\mathcal{R}_{k}^{\prime}}(\tau_2)} \, .\\
\end{eqnarray}

After summing over all contractions, the two‑point function corrected by tadpole insertions can be expressed as

\begin{eqnarray}\label{Two-Point-tadpole}
\langle  {\mathcal{R}}_{k_1}^{(2)} ({\tau _0}) \, {\mathcal{R}}_{k_2}^{(2)} ({\tau _0}) \rangle_{\Omega_{\text{tad}} }
= M_P^4 \int_{\tau_i}^{\tau_e} d \tau_1 \int_{\tau_i}^{\tau_1} d \tau_2 
\int_0^\infty \frac{d^3 \mathbf{q}}{(2 \pi)^3} \;
\mathcal{S}\!\left(\tau_1, \tau_2 ; q\right) ,
\end{eqnarray}

where the kernel $\mathcal{S}\!\left(\tau_1, \tau_2; q\right)$ is defined by the time integral

\begin{equation}\label{Tadpole-exp}
\mathcal{S}\!\left(\tau_1, \tau_2; q\right)  
= -4
\mathfrak{Im} \!\left[C^\ast({\tau _2}) B({\tau _1})\right] \;
\mathfrak{Im} \!\left[{\mathcal{X}}({\tau _1}) D({\tau _2})\right] .
\end{equation}

One possible realisation of the contractions in Eq.~\eqref{Tadpole-exp} is given by

\begin{eqnarray}\label{Tadpole-Ims00}
\mathcal{X}(\tau_1) &=&  
{\mathcal{R}_{k=0}(\tau)} \, {\mathcal{R}_{k=0}(\tau)} \,
{\mathcal{R}_{q}^{\ast}(\tau_1)} \, \mathcal{R}_{q}^{\ast\prime}(\tau_1) , \\
B(\tau_1) &=& \mathcal{R}_{q}^{\prime}(\tau_1) , 
\qquad C(\tau_2) = \mathcal{R}_{q}^{\ast}(\tau_2) , \qquad 
D(\tau_2) = \mathcal{R}_{k}^{\prime}(\tau_2) \, \mathcal{R}_{k}^{\ast\prime}(\tau_2) .
\end{eqnarray}

Alternatively, another valid contraction reads

\begin{eqnarray}\label{Tadpole-Ims01}
\mathcal{X}(\tau_1) &=&  
{\mathcal{R}_{k=0}(\tau)} \, {\mathcal{R}_{k=0}(\tau)} \,
{\mathcal{R}_{q}^{\ast\prime}(\tau_1)} \, \mathcal{R}_{q}^{\ast\prime}(\tau_1) , \\
B(\tau_1) &=& \mathcal{R}_{q}(\tau_1) , 
\qquad C(\tau_2) = \mathcal{R}_{q}^{\ast\prime}(\tau_2) , \qquad 
D(\tau_2) = \mathcal{R}_{k}(\tau_2) \, \mathcal{R}_{k}^{\ast\prime}(\tau_2) .
\end{eqnarray}

In practice, all such possible contractions must be taken into account when computing the tadpole contribution.

\subsection{Regularized tadpole contributions}

For the  bulk term  of the tadpole contribution we obtain

\begin{eqnarray}
		\label{Two-Point-TadpoleBulk-Result}\nonumber
	\left\langle\mathcal{R}_{k_1}^{(2)}\left(\tau_0\right) \mathcal{R}_{k_2}^{(2)}\left(\tau_0\right)\right\rangle_{\Omega_{\text {tad-bu }}}&=& \frac{3 e^{6 \Delta N} H^4}{32 M_p^4k_{\mathrm{CMB}}^3 \epsilon_i^2}\\\nonumber
\Bigg[\Big(3 \ln \left(-\Lambda \tau_i\right)&+&3(-2+3 \gamma_{EM}+\ln 8)\Big)(\cos 1-\sin 1)
	 +16 \pi(-1+\cos 1+\sin 1)\Bigg]\\
\end{eqnarray}

The gradient contribution turns out to be subleading, scaling as $e^{4 \Delta \mathcal{N}}$ rather than $e^{6 \Delta \mathcal{N}}$ :

\begin{equation}
	\left\langle\mathcal{R}_{k_1}^{(2)}\left(\tau_0\right) \mathcal{R}_{k_2}^{(2)}\left(\tau_0\right)\right\rangle_{\Omega_{\mathrm{tad}-\mathrm{gr}}}=\frac{81 H^4 e^{4 \Delta \mathcal{N}}}{256 M_P^4 \epsilon_i^2 k_{\mathrm{CMB}}^3} \delta_{k_1 k_2}\left[6 \gamma_{\mathrm{EM}}-11\right](\cos 1-\sin 1) .
\end{equation}

\subsection{Fractional correction from the tadpole}

The bulk contribution Eq.~\eqref{Two-Point-TadpoleBulk-Result} can be rewritten in terms of the power spectrum at CMB scales as
\begin{PSresult}[title={Fractional tadpole correction}]
\begin{equation}
	\begin{split}
\label{cubic-tadpole-cmb}
\frac{\Delta \mathcal{P}}{\mathcal{P}_\mathrm{CMB}}
= \frac{3\mathcal{P}_\mathrm{CMB}e^{6\Delta N}}{2 k_\mathrm{CMB}^3} &\Bigg[\Big(3 \ln \left(-\Lambda \tau_i\right)+3(-2+3 \gamma+\ln 8)\Big)(\cos 1-\sin 1) \\& +16 \pi(-1+\cos 1+\sin 1)\Bigg]
\end{split}
\end{equation}
\end{PSresult}
We see that this correction exhibits the same functional dependence on $\Delta\mathcal{N}$ and the same exponential enhancement $e^{6\Delta\mathcal{N}}$ as the cubic contribution derived earlier in Eq.~\eqref{fraction-cubic}. The numerical coefficient differs, but the overall scaling is identical, confirming that the tadpole diagram cannot be neglected in the total one‑loop correction.

\section{Renormalization of Infinities}
\label{sec:renorm}

In this section we present a systematic treatment of the various types of infinities that appear in our one-loop calculation. Since the divergences are of different origin, time-like, ultraviolet power-law, and logarithmic, we employ distinct regularization and renormalization strategies. Time divergences are handled via the \(i\epsilon\) prescription and the Cauchy principal value, UV power-law divergences are removed by introducing appropriate counterterms, and logarithmic divergences are resummed using renormalization group (RG) flow techniques. Rather than listing every such divergence, we provide representative examples of each type and demonstrate the corresponding cancellation mechanism.

\subsection{Classification of Divergences}
\label{classify-divergs}

We begin by illustrating the structure of time divergences that arise from the time-time part,  for instance of the cubic Hamiltonian \(H_3\). Table~\ref{Timedives0} collects the coefficients of inverse powers of {\(s \equiv \tau_1 - \tau_i\)} and \(r \equiv \tau_2 - \tau_1\), as well as the logarithmic contributions, after expansion in the parameters \(\tau_e\) and \(\tau_i\).

\begin{table}[h]
\centering
\begin{tabular}{l l}
\toprule
\textbf{Term} & \textbf{Expression} \\
\midrule
\multicolumn{2}{c}{\textbf{Inverse powers of \(s={(\tau_1-\tau_i)}\)}} \\
\midrule
\(1/s\) & \(-\dfrac{45  H^2 \tau_i^6}{32 M_p^2\tau_1^2 \tau_e^3 \epsilon_i}\) \\
\\
\(1/s^2\) & \(\dfrac{9  H^2 \tau_i^4 (3 \tau_1^2 - 10 \tau_i^2)}{64 M_p^2\tau_1 \tau_e^3 \epsilon_i}\) \\
\(1/s^3\) & \(-\dfrac{3  H^2 \tau_i^6}{32 M_p^2\tau_1^3 \epsilon_i}\) \\
\(1/s^4\) & \(-\dfrac{9  H^2 \tau_1 \tau_i^6}{64 M_p^2\tau_e^3 \epsilon_i}\) \\
\midrule
\multicolumn{2}{c}{\textbf{Inverse powers of \(r=\tau_2-\tau_1\)}} \\
\midrule
\(1/r\) & \(-\dfrac{9  H^2 \tau_i^3 (8 \tau_1^2 - 3 \tau_i^2)}{16 M_p^2\tau_1 \tau_e^3 \epsilon_i}\) \\
\(1/r^2\) & \(\dfrac{45  H^2 \tau_1 \tau_i^4}{64 M_p^2\tau_e^3 \epsilon_i}\) \\
\(1/r^3\) & \(\dfrac{9  H^2 \tau_1 \tau_i^5}{32 M_p^2\tau_e^3 \epsilon_i}\) \\
\midrule
\multicolumn{2}{c}{\textbf{Logarithmic terms}} \\
\midrule
\(\log\!\left(\dfrac{s \tau_i}{\tau_1^2 - \tau_1 \tau_i}\right)\) & \(-\dfrac{81 i H^2 \tau_i^4}{2 M_p^2\tau_1 \tau_e^3 \epsilon_i} - \dfrac{9 i H^2 \tau_i^6}{2 M_p^2\tau_1^3 \tau_e^3 \epsilon_i}\) \\
\\
\(\log\!\left(\dfrac{r \tau_1}{(\tau_1 - \tau_i) \tau_i}\right)\) & \(\dfrac{81 i H^2 \tau_1 \tau_i^2}{2 M_p^2\tau_e^3 \epsilon_i} - \dfrac{81 i H^2 \tau_i^4}{4 M_p^2\tau_1 \tau_e^3 \epsilon_i}\) \\
\\
\(\log\!\left(\dfrac{r}{\tau_1 - \tau_i}\right)\) & \(\dfrac{189 i H^2 \tau_i^4}{16 M_p^2\tau_1 \tau_e^3 \epsilon_i}\) \\
\\
\(\log\!\left(\dfrac{s}{\tau_1 - \tau_i}\right)\) & \(\dfrac{27 i H^2 \tau_1 \tau_i^2}{8 M_p^2 \tau_e^3 \epsilon_i} - \dfrac{27 i H^2 \tau_i^3}{8 M_p^2\tau_e^3 \epsilon_i} + \dfrac{405i H^2 \tau_i^4}{16 M_p^2\tau_1 \tau_e^3 \epsilon_i} - \dfrac{9  H^2 \tau_i^6}{8 M_p^2\tau_1^3 \tau_e^3 \epsilon_i}\) \\
\bottomrule
\end{tabular}
\caption{Coefficients of the inverse power and logarithmic divergences after expansion in \(\tau_e\) and \(\tau_i\).}
\label{Timedives0}
\end{table}

We now detail the cancellation of these time divergences. Consider, for example, the combination of the first four rows of Table~\ref{Timedives0} together with the logarithmic terms. The coefficients of the \(1/s\) and \(1/s^3\) terms are positive, while those of \(1/s^2\) and \(1/s^4\) are negative. By applying the Cauchy principal value prescription and appropriately tuning the rate at which the divergent terms tend to infinity, we can achieve mutual cancellation. Specifically, we write

\begin{equation}
	\begin{split}
	&-\dfrac{45  H^2 \tau_i^6}{32 M_p^2 s \tau_1^2 \tau_e^3 \epsilon_i}+\dfrac{9  H^2 \tau_i^4 (3 \tau_1^2 - 10 \tau_i^2)}{64 M_p^2 s^2 \tau_1 \tau_e^3 \epsilon_i}+\\
	&-\dfrac{3  H^2 \tau_i^6}{32 M_p^2 s^3 \tau_1^3 \epsilon_i}+(-\dfrac{81 i H^2 \tau_i^4}{2 M_p^2\tau_1 \tau_e^3 \epsilon_i} - \dfrac{9 i H^2 \tau_i^6}{2 M_p^2\tau_1^3 \tau_e^3 \epsilon_i})\log\!\left(\dfrac{s \tau_i}{\tau_1^2 - \tau_1 \tau_i}\right)+\\&\log\!\left(\dfrac{s}{\tau_1 - \tau_i}\right)\Big(\dfrac{27 i H^2 \tau_1 \tau_i^2}{8 M_p^2\tau_e^3 \epsilon_i} - \dfrac{27 i H^2 \tau_i^3}{8 M_p^2\tau_e^3 \epsilon_i} + \dfrac{405 i H^2 \tau_i^4}{16M_p^2 \tau_1 \tau_e^3 \epsilon_i} - \dfrac{9 i H^2 \tau_i^6}{8 M_p^2\tau_1^3 \tau_e^3 \epsilon_i}\Big)-\dfrac{9  H^2 \tau_1 \tau_i^6}{64 M_p^2 \tilde{s}^4\tau_e^3 \epsilon_i}=0.
	\end{split}
\end{equation}

By choosing the parameter \(\tilde{s}\) appropriately, the above combination can be made to vanish. When the signs of the terms are uniform, one may use the replacement \(r \to r + i\epsilon\) (with \(\epsilon \to 0\)) to flip the sign of a given term, thereby enabling cancellation. In this manner, all time divergences of this type are consistently removed.

For oscillating infinities—which arise, for instance, from the quartic Hamiltonian \(H_4\), we adopt a different regularization. Table~\ref{Oscil0} displays two representative terms.

\begin{table}[h]
\centering
\begin{tabular}{l l}
\toprule
\textbf{Term} & \textbf{Expression} \\
\midrule
Oscillating term 1 & 
$\displaystyle \frac{27 H^4}{32 M_p^4 k  \epsilon_i^2} \left( \frac{1}{\tau_e} + \frac{1}{ \tau_i } \right) e^{2 i \Lambda (\tau_e - \tau_i)}$\\
\\
Oscillating term 2 & 
$\displaystyle   \frac{63 i H^4}{128 M_p^4\epsilon_i^2} \left(3 + 4\frac{\tau_e}{\tau_i} + 2\frac{\tau_i}{\tau_e} + \frac{\tau_i^2}{\tau_e^2} \right) e^{-2 i \Lambda (\tau_e - \tau_i)}$ \\
\bottomrule
\end{tabular}
\caption{Oscillating divergent terms with exponential phase factors.}
\label{Oscil0}
\end{table}

To remove these oscillating divergences, we deform the upper limit of the momentum integral as \(\Lambda \to \Lambda (1 \pm i \epsilon)\), which effectively damps the exponential factors and renders the integrals finite.

Finally, we exhibit a typical UV power-law divergence originating from the \(H_4\) contribution, as shown in Table~\ref{Momentum-Inf}. These divergences will be cancelled by the counterterms introduced in the next subsection.

\begin{table}[h]
\centering
\begin{tabular}{l}
\toprule
\textbf{Momentum infinities ($H_4$ contribution)} \\
\midrule
$\displaystyle 
~~~~~~~~~~~~~\frac{{{H^4}}}{{128 {\epsilon_i^2M_p^4}{\pi ^2}}}[\underbrace {21{\Lambda ^2}\tau _i^2\frac{{\tau _i^4}}{{\tau _e^4}}}_{{\Lambda ^2}} + \underbrace {12{\Lambda ^4}\tau _i^4\frac{{\tau _i^3}}{{\tau _e^3}}}_{{\Lambda ^4}}]
$ \\
\bottomrule
\end{tabular}
\caption{Momentum power-law divergences from the $H_4$ diagram.}
\label{Momentum-Inf}
\end{table}

The counterterms required to absorb these divergences are listed in Table~\ref{CounterTerms}. The first row corresponds to the \(\Lambda^2\) and \(\Lambda^4\) divergences, while the second row provides the counterterm for the bare mass, where \(\mathbb{T}\) denotes the time unit.

\begin{table}[h]
\centering
\begin{tabular}{l l}
\toprule
\textbf{Term} & \textbf{Expression} \\
\midrule
Counter term of $\frac{H^3}{M_p^3}$ & 
$\displaystyle -\frac{i e^{3 \Delta N} \frac{H^3}{M_p^3} \Lambda^4 \tau_i^4}{9 \pi^2 \Delta N}$ \\
\\
Counter term  $\frac{m^2}{M_p^2}$& 
$\displaystyle -\frac{i e^{3 \Delta N} \frac{H^2}{M_p^2}  \Lambda^4 t_i^4}{3 \pi^2 \log(- \tau_0 / \mathbb{T}) \, \Delta N}$ \\
\bottomrule
\end{tabular}
\caption{Counterterms for the UV divergences: the first row cancels the \(\Lambda^2\) and \(\Lambda^4\) terms, while the second row renormalizes the bare mass. Here \(\mathrm{T}\) is the time unit.}
\label{CounterTerms}
\end{table}

\subsection{Counterterms and UV Renormalization}

\begin{figure}[ht]
    \centering
    \includegraphics[width=0.88\linewidth]{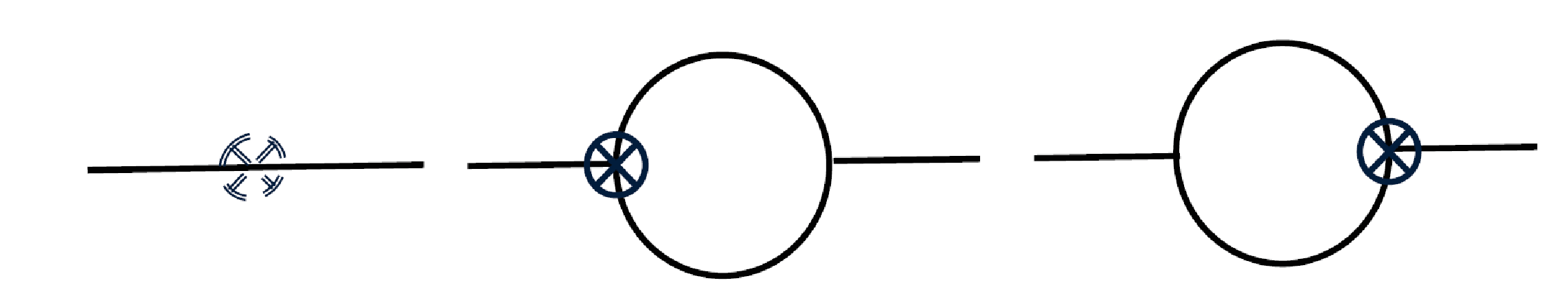}
    \caption{Diagrammatic representation of the counterterm cancellations for the UV divergences.}
    \label{fig2}
\end{figure}

We now discuss the cancellation of UV divergences through the redefinition of the mass and the cubic coupling in the Hamiltonian. The procedure is schematically depicted in Fig.~\ref{fig2}. In essence, we introduce a bare mass term that absorbs the divergent contributions, closely following the standard P\&S renormalization scheme.

Following Ref.~\cite{Chen:2009zp}, we write the Mukhanov–Sasaki equation for a massive field as
\begin{equation}\label{MS-chen}
  \mathcal{R}_k^{\prime \prime} - \frac{2}{\tau} \mathcal{R}_k^{\prime} + k^{2} \mathcal{R}_k + \frac{m^{2}}{H^{2}\tau^{2}} \mathcal{R}_k =  0,
\end{equation}
whose solution for \(m^{2}/H^{2} \leq 9/4\) is given by
\begin{equation}\label{Hankel0}
\mathcal{R}_k = -i e^{i(\nu +\frac{1}{2})\frac{\pi}{2}} \frac{\sqrt{\pi}}{2} H (-\tau)^{3/2} H_{\nu}^{(1)}(-k\tau),\quad \nu=\sqrt{9/4-m^{2}/H^{2}}.
\end{equation}
Expanding this solution in the small-mass limit and taking the superhorizon limit yields
\begin{equation}\label{seriesmasssmall}
\left<\mathcal{R}_{k_1}\mathcal{R}_{k_2}\right>=\frac{H^2+\frac{4m^2}{3}\big(-2+\gamma_{EM}+ \log(-8 k \tau)\big)}{4 \epsilon_i(M_P^2) k_1^3}(2\pi)^3\delta^3(k_1+k_2),
\end{equation}
which we adopt as our renormalized propagator.

Including the mass term also modifies the zero-mode solution. The general solution for the zero mode takes the form
\begin{equation}
	\mathcal{R}_{k=0}=\tilde{c}_1\left(\frac{\tau}{\tau_*}\right)^{\frac{3}{2}(1+\sqrt{1-\frac{4 m^2}{9H^2}})}+\tilde{c}_2\left(\frac{\tau}{\tau_*}\right)^{\frac{3}{2}(1-\sqrt{1-\frac{4 m^2}{9H^2}})},
\end{equation}
where \(\tau_*\) is an arbitrary reference time. Expanding in \(m^2/H^2\) and applying the matching conditions detailed in Appendix~\ref{AppC}, we obtain the coefficients at the end of inflation:
\begin{equation}
\tilde{c}_1 + \frac{\tau^3 \, \tilde{c}_2}{3} + \frac{m^2 \left( \tilde{c}_1 - \frac{\tau^3 \, \tilde{c}_2}{3} \right) \left( i \pi + \log\big(-\frac{\tau_0}{\tau_*}\big) \right)}{3 H^2},
\end{equation}
with
\begin{equation}
	\tilde{c}_1=\frac{H}{2 M_p \sqrt{k_{\text{CMB}}^3 \epsilon_i}} - \frac{i m^2 \left( \pi - i \log\big(-\frac{\tau_0}{\tau_*}\big) \right)}{6 H M_p \sqrt{k_{\text{CMB}}^3 \epsilon_i}},
\end{equation}
\begin{equation}
 \tilde{c}_2=-\frac{m^2}{6 H M_p\tau_e^3 \sqrt{k_{\text{CMB}}^3 \epsilon_i}} + \frac{3 \left( 1 - (1 - i) e^i \right) H k_{\text{CMB}}^3 \tau_i^6}{2 M_p\tau_e^6 \sqrt{k_{\text{CMB}}^3 \epsilon_i}}.
\end{equation}
Here \(\tau_0\) denotes the time at the end of inflation.

\subsection{Callan–Symanzik Equation and Renormalization Group Flow}
\label{sec:CS_RG_flow}

We now apply the Callan–Symanzik formalism, as reviewed in Sec.~\ref{sec:RG_review} (see Eq.~\eqref{eq:CS_NLSM}), to derive the running of the couplings. As a concrete illustration, we solve the CS equation to obtain the anomalous dimension \(\gamma\) and the beta function \(\beta\).

To facilitate comparison with the standard P\&S RG procedure, we introduce the exponential form for the curvature perturbation:
\[
e^{\mathcal{R}} = 1 + \mathcal{R} + \frac{\mathcal{R}^2}{2} + \cdots .
\]
We focus on the logarithmic terms that scale as \(1/p^3\), which appear in both the quartic (\(H_4\)) and tadpole contributions.

\subsubsection{One-Point Function and the Anomalous Dimension}

The one-point function of \(e^{\mathcal{R}}\) is expanded as
\begin{equation}
	\left<(e^{\mathcal{R}})_{p}\right>=\delta^3(0)+\left<\mathcal{R}_p\right>+\frac{\left<\mathcal{R}^2\right>_{p}}{2}+\mathcal{O}(\mathcal{R}^3).
\end{equation}
To leading order,
\begin{equation}
	\label{H3one}
\left<\mathcal{R}_p\right>=2 \mathfrak{Im}\left(\left<\mathcal{R}_p\int H_3dt\right>\right),
\end{equation}
where the contraction of fields shows that only the zero mode contributes. Similarly,
\begin{equation}
	\left<(\mathcal{R}^2)_p\right>=\left<\int \frac{d^3 q}{(2\pi)^3}\mathcal{R}_{p-q}\mathcal{R}_q\right>.
\end{equation}

After explicit computation, the one-point function reads
\begin{equation}\label{one-point}
\begin{aligned}
\langle (e^\mathcal{R})_p\rangle =\Bigg( \frac{H^2}{M_p^2\epsilon_i} \biggl[
&\frac{\tau_i^6 \log(-\Lambda \tau_i)}{8\,\tau_e^6} \\
&+ \bigl(\cos 1 - \sin 1\bigr) \Biggl(
   \frac{9\,\tau_i^2 \bigl(12 \tau_e^4 + 9 \tau_e^2 t_i^2 + 4 \tau_i^4\bigr) \log\!\bigl(2\Lambda (\tau_e - \tau_i)\bigr)}{64\,\tau_e^6} \\
&\qquad + \frac{3 \Lambda^4 \tau_i^6}{64\,\tau_e^2} + \frac{9 \Lambda^2 \tau_i^6}{32\,\tau_e^4} \Biggr)
\biggr] + \text{finite terms}\Bigg)(2\pi)^3\delta^3(p).
\end{aligned}
\end{equation}
The delta function confirms that only the zero mode (\(p=0\)) contributes, analogous to the sigma-field one-point function in Eq.~\eqref{sigma-one}.

To remove the power-law divergences, we introduce counterterms for both the one-point and two-point functions. According to Eq.~\eqref{H3one}, we need counterterms in the cubic Hamiltonian. Defining \(\mathcal{H} \equiv H^3/M_p^3\), we set bare parameters as \(\mathcal{H}_B = \mathcal{H} + \delta\mathcal{H}\) and \(m_B^2 = m^2 + \delta m^2\). The conditions for cancellation of \(\Lambda^4\) divergences are

\begin{equation}
\begin{split}
\frac{\delta H }{(\frac{H}{M_P})^3}\left<	\mathcal{R}_p\mathcal{R}_p\right>_{H_3}+\left<	\mathcal{R}_p\mathcal{R}_p\right>_{\mathbb{INF},H_4}+\frac{\delta m^2\big(-2+\gamma_{EM}+ \log(-8 k \tau_0)\big)}{3 \epsilon_i(M_P^2) k^3}=0,
\end{split}
\end{equation}
and
\begin{equation}
	\begin{split}
		\frac{\delta H }{(\frac{H}{M_P})^3}\left<	(e^{\mathcal{R}})_p\right>_{\mathbb{INF}}+\left	<(e^{\mathcal{R}})_p\right>_{\mathbb{INF}}+\frac{\delta m^2 \tau_i^4 e^{2\Delta N}}{64 \epsilon_iM_p^2}(\log(-\frac{\tau_0}{\tau_*})(-\cos 1+\sin 1))\Lambda^4=0.
	\end{split}
\end{equation}
Here the subscript \(\mathbb{INF}\) denotes the \(\Lambda^4\)-proportional divergent parts (whose explicit coefficients are prolong and omitted for brevity). Solving these two equations simultaneously yields the counterterms displayed in Table~\ref{CounterTerms}, where we have expanded the final results in the limit \(\log(-\tau_0/\mathbb{T}) \to -\infty\). The corresponding diagrammatic representation appears in Fig.~\ref{fig2}.

We now turn to the logarithmic divergences. These appear as terms like \(\log(-\Lambda \tau_i)\) in both the one-point and two-point functions. We define the renormalization scale in terms of time rather than momentum: the scale is set at the transition time \(\tau_i'\) via
\begin{equation}
	\left<\mathcal{R}_p\mathcal{R}_p\right>|_{\tau_i=\tau_i'}=\frac{\mathcal{P}_{\text{CMB}}}{k_{\text{CMB}}^3},
\end{equation}
and
\begin{equation}
	\left<(e^\mathcal{R})_p\right>|_{\tau_i=\tau_i'}=\delta^3(0).
\end{equation}
Due to quantic nature of perturbations and their backreaction, the transition time may differ for different patches.
Physically, this means that the two-point function assumes its standard value at a patch where the transition occurs at \(\tau_i'\), while the one-point function is normalized to zero at the same patch. The transition time corresponds to the mode crossing the horizon at that instant:
\begin{equation}
	\label{ptotime}
	M=\frac{-1}{\tau_i'}.
\end{equation}
Since we are interested in superhorizon scales that exit at the beginning of inflation, we take \(M \simeq k_{\text{CMB}}\). The logarithmic counterterms for the RG flow are then expressed as
\begin{equation}
	\label{onelog}
\left<(e^\mathcal{R})_p\right>'=\frac{(9(\cos 1- \sin 1)+2) e^{6\Delta N} H^2}{16 \epsilon_i M_p^2}\left(\log(-\Lambda \tau_i)-\log(-\Lambda \tau_i')\right)+\text{finite terms},
\end{equation}
where the prime indicates that the delta function has been removed. Similarly, the two-point function becomes
\begin{equation}
	\left<\mathcal{R}_p\mathcal{R}_p\right>'=-\mathcal{P}_{\text{CMB}}^2\frac{27 e^{6 \Delta N}}{2}\left(\log(-\Lambda \tau_i)-\log(-\Lambda \tau'_i)\right)+\text{finite terms}.
\end{equation}

Following the iterative P\&S procedure, we use Eq.~\eqref{eq:CS_NLSM} for \(n=1\) and \(n=2\). Since the beta function is of higher order in the coupling than the anomalous dimension, we neglect \(\beta\) in the \(n=1\) equation. Using \eqref{ptotime}, the leading-order CS equation reads
\begin{equation}
	\label{CS1}
\left(M\frac{\partial}{\partial M}+\gamma(\lambda)\right)\left<(e^\mathcal{R})_p\right>=0.
\end{equation}
From this we extract
\begin{equation}
\gamma(\lambda) = \frac{H^2 \tau_i^6 \bigl(9(\sin 1 - \cos 1) - 2\bigr)}{16M-P^2\,\tau_e^6 \,\epsilon_i}.
\end{equation}

\subsubsection{Two-Point Function and the Beta Function}

Armed with \(\gamma(\lambda)\), we now determine the beta function from the CS equation for \(n=2\):
\begin{equation}
	\left(M \frac{\partial}{\partial M}+\beta\frac{\partial}{\partial\lambda}+\gamma\right)\left<\mathcal{R}_p\mathcal{R}_p\right>'=0.
\end{equation}
The solution yields
\begin{equation}\label{betamain}
\beta(\lambda, M) = \frac{9 H^3 \tau_i^6 \bigl(110 + 9(\cos 1 - \sin 1)\bigr)}{16 M_P^3 \tau_e^6 \,\epsilon_i}.
\end{equation}

\subsubsection{Running Coupling Constant}

We define \(\bar{\lambda} \equiv \bar{H}/M_p\) as the running coupling and \(\lambda \equiv H/M_p\) at a reference scale. The RG equation for the running coupling is
\begin{equation}\label{runningcoupling}
\frac{d}{d\log(p/M)}\,\bar{\lambda}(p;\lambda) = \beta(\bar{\lambda}), \qquad \bar{\lambda}(M;\lambda) = \lambda.
\end{equation}
Inserting the beta function from Eq.~\eqref{betamain} and integrating gives
\begin{equation}\label{lambdabar}
\bar{\lambda} = \frac{H^2}{M_P^2 \left(1 - \dfrac{H^2 e^{6\Delta N} \log(p/M) \bigl(110+ 9(\cos 1 - \sin 1)\bigr)}{8 M_P^2  \,\epsilon_i}\right)}.
\end{equation}

This result exhibits the expected qualitative behavior of RG flow: the coupling runs logarithmically with momentum \(p\), and the running is suppressed by the exponential factor \(e^{6\Delta\mathcal{N}}\) characteristic of the USR phase. This confirms the consistency of our renormalization procedure with standard renormalization group analysis.

\newpage

\section{Conclusion}
\label{conclusion0}

In the ongoing debate concerning one-loop corrections in models of ultra-slow-roll (USR) inflation, three distinct viewpoints have emerged in the literature. One perspective holds that the fractional loop corrections scale with the peak of the power spectrum at the end of the USR phase, so that $\Delta\mathcal{P}/\mathcal{P} \sim \mathcal{P}_{\mathrm{peak}} \sim e^{6\Delta\mathcal{N}} \mathcal{P}_{\mathrm{CMB}}$ \cite{Kristiano:2022maq, Firouzjahi:2023aum, Firouzjahi:2023bkt}. A second viewpoint argues that these corrections are suppressed by the slow-roll parameter $\epsilon$ in the final phase, and therefore can remain perturbative if the USR–SRII transition is sufficiently smooth \cite{Riotto:2023hoz, Riotto:2023gpm, Iacconi:2023ggt}. A third proposal maintains that the loop corrections are nearly negligible due to volume suppression or other cancellation mechanisms \cite{Fumagalli:2023hpa, Tada:2023rgp}.\\
In this work we have performed a systematic one-loop calculation in the three-phase SRI–USR–SRII model. Our analysis includes all relevant diagrams, cubic, quartic, and tadpole, and employs a rigorous UV-IR cutoff regularization and adequate renormalization schemes. To correctly capture the contribution of all momentum scales, we integrate the loop momentum over the entire range from $0$ to $\infty$, rather than restricting to a limited interval as done in some earlier studies.\\
 Time integrals are regularized using the Cauchy principal value and the $i\epsilon$ prescription, which effectively isolates the finite, physical parts of nested time integrations. The power law forms of ultraviolet divergences that appear are systematically removed by introducing appropriate counterterms, following the standard Peskin–Schroeder renormalization procedure.\\
Beyond the subtraction of divergences, a central novelty of our work is the explicit renormalization group (RG) improvement of the loop corrections. We have made extensive use of the Callan–Symanzik (CS) equation to derive the running of the couplings. To this end, we introduced a mass parameter for the curvature perturbation and computed the renormalized propagator in the superhorizon limit. We then calculated the one-point function of $e^{\mathcal{R}}$ and used it to extract the running parameter $\gamma(\lambda),$ as
\begin{equation}
\gamma(\lambda) = \frac{H^2 \tau_i^6 \bigl(9(\sin 1 - \cos 1) - 2\bigr)}{16 M_P^2\,\tau_e^6 \,\epsilon_i}.
\label{eq:gamma_final}
\end{equation}
This represents a key result of our renormalization procedure: the anomalous dimension encodes the scale dependence of the field strength and is directly linked to the exponential enhancement characteristic of the USR phase. The presence of the factor $\tau_i^6/\tau_e^6 = e^{6\Delta\mathcal{N}}$ in $\gamma$ confirms that the running of the field normalization is itself amplified by the non-attractor dynamics.
Furthermore, by applying the CS equation to the two-point function, we obtained the beta function
\begin{equation}
\beta(\lambda, M) = \frac{9 H^3 \tau_i^6 \bigl(110 + 9(\cos 1 - \sin 1)\bigr)}{16 M_P^3 \tau_e^6 \,\epsilon_i},
\label{eq:beta_final}
\end{equation}
which governs the running of the effective coupling $\lambda \equiv H/M_P$. Solving the renormalization group equation for the running coupling yields
\begin{equation}
\bar{\lambda} = \frac{H^2}{M_P^2 \left(1 - \dfrac{H^2 e^{6\Delta N} \log(p/M) \bigl(110+ 9(\cos 1 - \sin 1)\bigr)}{8 M_P^2  \,\epsilon_i}\right)}.
\label{eq:lambdabar_final}
\end{equation}

Our RG analysis reveals that the logarithmic divergences of the form $\log(-\Lambda\tau_i)$ are resummed into the running couplings, thereby rendering the final physical results independent of the arbitrary cutoff $\Lambda$. The renormalization scale is naturally identified with the comoving momentum scale $p$, or equivalently with the transition time via $M = -1/\tau_i'$. \\
This time–momentum duality is particularly natural in inflationary backgrounds, where mode crossing relates time scales to momentum scales. The renormalization conditions we imposed, fixing the two-point function and the one-point function at the reference patch, are physically motivated by the observation that the transition time can vary across different Hubble patches due to quantum perturbations.\\
After this careful renormalization and RG improvement, we find that the regularized one-loop corrections from all diagrams, cubic, quartic, and tadpole, consistently scale as
\begin{equation}
\frac{\Delta \mathcal{P}}{\mathcal{P}_\mathrm{CMB}} \sim \mathcal{P}_\mathrm{CMB} \, e^{6\Delta\mathcal{N}} \, \mathcal{F}(\Delta\mathcal{N}),
\end{equation}
where $\mathcal{F}(\Delta\mathcal{N})$ is a polynomial in $\Delta\mathcal{N}$ (with leading linear term from the cubic and tadpole contributions) that can further enhance the correction for $\Delta\mathcal{N}\gtrsim 1$. Numerical coefficients depend on the specific diagram, but the exponential enhancement is universal. For typical values $\Delta\mathcal{N} \sim 2$–$3$ and $\mathcal{P}_\mathrm{CMB} \sim 2\times10^{-9}$, the fractional correction can be of order $10^{-4}$–$10^{-1}$. Consequently, our analysis supports the conclusion that loop corrections can become non-perturbatively large if the USR phase is sufficiently long and the transitions are sharp, thereby confirming the first of the three viewpoints. This conclusion is robust because it arises from the renormalized, cutoff-independent result and is not an artifact of a particular regularization scheme.\\
Primordial black hole formation scenarios typically require $\mathcal{P}_{\mathrm{peak}} \sim 10^{-2}$–$10^{-1}$ on small scales. Our results imply that the very same enhancement that produces PBHs also amplifies loop corrections to CMB scales. For such scenarios to remain viable, the USR phase must be short enough ($\Delta\mathcal{N} \lesssim 2$) or the transition to SRII must be sufficiently mild to allow for slow-roll suppression. A more detailed exploration of the full parameter space, including arbitrary values of the transition sharpness parameter $h$, is left for future work.\\
As a non-perturbative approach, we have also extended the stochastic inflation formalism to incorporate interacting noise, which captures the nonlinear interactions characteristic of quantum loops \cite{Nassiri-Rad:2025dsa}. This generalized approach suggests that the stochastic picture can successfully encompass a broader class of quantum phenomena and may provide complementary insights into the behavior of loop corrections in multi-phase inflation.\\
Looking ahead, it would be natural to extend this investigation using alternative techniques, such as lattice field theory methods, or to consider higher-order effects (e.g., renormalized two-loop corrections \cite{Two-loop}) in order to better understand the behavior of loop corrections in multi-phase inflation. We intend to address these directions in future work.

\acknowledgments

We are grateful to Hassan Firouzjahi for carefully reviewing the initial draft of this paper and for providing constructive suggestions. We also thank Sina Hooshangi for the discussions on the early version of this work.

\appendix

\section{Quartic Hamiltonian Integrands and Coefficients}
\label{AppA}

In this appendix we present the explicit expressions for the quartic Hamiltonian integrands that arise during the USR phase. The three distinct contributions are

\begin{eqnarray}\label{I-one-three-1}
  I_1 &=& \tilde{I}\,{\left| \mathcal{R}_\mathbf{q}^\prime (\tau) \right|^2}
  \mathfrak{Im} \left[ \mathcal{R}_{\mathbf{k}}^*(\tau_0)^2 \mathcal{R}_{\mathbf{k}}(\tau)^2 \right] \,, \\[4pt]
\label{I-one-three-2}
  I_2 &=& \tilde{I}\, \mathfrak{Im} \left[ \mathcal{R}_{\mathbf{k}}^*(\tau_0)^2 \mathcal{R}_{\mathbf{k}}(\tau) \mathcal{R}_{\mathbf{k}}^\prime(\tau) \mathcal{R}_{\mathbf{q}}(\tau) \mathcal{R}_{\mathbf{q}}^\prime(\tau)^* \right] \,, \\[4pt]
\label{I-one-three-3}
  I_3 &=& \bar{I}\, q^2 {\left| \mathcal{R}_\mathbf{q}(\tau) \right|^2}
  \operatorname{Im} \left[ \mathcal{R}_{\mathbf{k}}^*(\tau_0)^2 \mathcal{R}_{\mathbf{k}}(\tau)^2 \right] \,,
\end{eqnarray}

where we have introduced the shorthand notations

\[
\tilde{I} \equiv \left( \eta^2 - \frac{\eta^\prime}{aH} \right) a^2(\tau), \qquad
\bar{I} \equiv \left( \eta^2 + \frac{\eta^\prime}{aH} \right) a^2(\tau).
\]

The Bogoliubov coefficients appearing in the SRII mode function, Eq.~\eqref{SRIIMode}, are obtained from the matching conditions at the transition surfaces between the phases. After a lengthy but straightforward calculation, they are given by

\begin{PSresult}[title={Bogoliubov coefficients for SRII}]
\begin{eqnarray}\label{alpha-beta3}
\alpha_k &=& \frac{1}{8 k^6 \tau_i^3 \tau_e^3} \Big[ 3h (1 - i k \tau_e)^2 (1 + i k \tau_i)^2 e^{2i k (\tau_e - \tau_i)} \nonumber\\
&\phantom{=}& - i (2 k^3 \tau_i^3 + 3i k^2 \tau_i^2 + 3i) (4 i k^3 \tau_e^3 - h k^2 \tau_e^2 - h) \Big], \\[6pt]
\label{beta-beta3}
\beta_k &=& \frac{-1}{8 k^6 \tau_i^3 \tau_e^3}
 \Big[ 3 (1 + i k \tau_i)^2 (h + h k^2 \tau_e^2 + 4 i k^3 \tau_e^3) e^{-2i k \tau_i} \nonumber\\
&\phantom{=}& + i h (1 + i k \tau_e)^2 (3i + 3i k^2 \tau_i^2 + 2 k^3 \tau_i^3) e^{-2i k \tau_e} \Big] \,.
\end{eqnarray}
\end{PSresult}

The corresponding complex conjugates $\alpha_k^*$ and $\beta_k^*$ can be derived analogously.

\section{Cubic Hamiltonian Integrands and Coefficients}
\label{AppB}

For the cubic Hamiltonian, the central quantity is the function $\mathcal{G}(\tau_1,\tau_2;q)$ defined in Eq.~\eqref{Two-Point-Cubic01}. Its explicit form is

\begin{eqnarray}
\mathcal{G}\left( \tau_1,\tau_2;q \right) &\equiv& \mathfrak{Im}\big[ G^*(\tau_2) g_k^*(\tau_2) \big( Z(\tau_1) + 2Y(\tau_1) \big) \big] \label{G-cal1}\\
&-& \mathfrak{Im}\Big[ G^*(\tau_2) \big[ \big( Z(\tau_1) + 2Y(\tau_1) \big) + g_k(\tau_1) Z(\tau_1) \big] \Big] \label{G-cal2}\\
&+& \mathfrak{Im} \big[ \tilde{G}^*(\tau_2) \tilde{Z}(\tau_1) \big] \label{G-cal3}
\end{eqnarray}

where the building blocks are defined as

\begin{eqnarray}\label{G-curly}
G^*(\tau_2) &=& \eta \epsilon a^2 \mathcal{R}_{\mathbf{k}}^*(\tau_0) \mathcal{R}_{\mathbf{k}}(\tau_2) \mathcal{R}_{\mathbf{q}}^{\prime}(\tau_2)^2, \\[4pt]
\tilde{G}^*(\tau_2) &=& \mathbf{q}^2 \eta \epsilon a^2 \mathcal{R}_{\mathbf{k}}^*(\tau_0) \mathcal{R}_{\mathbf{k}}(\tau_2) \mathcal{R}_{\mathbf{q}}(\tau_2)^2, \\[4pt]
Z(\tau_1) &=& 2 \epsilon \eta a^2 \mathcal{R}_{\mathbf{q}}^{\prime}(\tau_1)^2 \mathfrak{Im}\left[ \mathcal{R}_{\mathbf{k}}^*(\tau_0) \mathcal{R}_{\mathbf{k}}(\tau_1) \right], \\[4pt]
\tilde{Z}(\tau_1) &=& 2 \mathbf{q}^2 \epsilon \eta a^2 \mathcal{R}_{\mathbf{q}}(\tau_1)^2 \mathfrak{Im}\left[ \mathcal{R}_{\mathbf{k}}^*(\tau_0) \mathcal{R}_{\mathbf{k}}(\tau_1) \right], \\[4pt]
Y(\tau_1) &=& 2 \epsilon \eta a^2 \mathcal{R}_{\mathbf{q}}^{\prime}(\tau_1) \mathcal{R}_{\mathbf{q}}(\tau_1) \mathfrak{Im}\left[ \mathcal{R}_{\mathbf{k}}^*(\tau_0) \mathcal{R}_{\mathbf{k}}^{\prime}(\tau_1) \right], \\[4pt]
g_{\mathbf{q}}(\tau) &=& -\frac{(\partial \mathcal{R}_{\mathbf{q}})^2}{\mathcal{R}_{\mathbf{q}}^{\prime 2}}
= -\frac{q^2 \mathcal{R}_{\mathbf{q}}^2}{\mathcal{R}_{\mathbf{q}}^{\prime 2}}.
\end{eqnarray}

\section{Tadpole Calculations}
\label{AppC}

Finally, we address the tadpole contributions corresponding to the zero‑mode of the curvature perturbation. Since the relevant vertices are cubic (see $\mathbf{H}_3$), the tadpole diagrams require solving the Mukhanov–Sasaki equation in the limit $k=0$ during each phase.

\subsection{Zero‑mode solutions}

By solving Eq.~\eqref{M-K} with $k=0$, we obtain the following zero‑mode solutions:

\begin{PSresult}[title={Zero‑mode solutions for each phase}]
\begin{itemize}
\item SRI phase:
\begin{equation}\label{ZeromodeSRI}
\mathcal{Q}_{k=0}^{(1)} = \frac{c_1}{3} \tau^3 + c_2.
\end{equation}

\item USR phase:
\begin{equation}\label{ZeromodeUSR}
\mathcal{Q}_{k=0}^{(2)} = \frac{d_1}{3} \tau^{-3} + d_2.
\end{equation}

\item SRII phase:
\begin{equation}\label{ZeromodeSRII}
\mathcal{Q}_{k=0}^{(3)} = \frac{e_1}{3} \tau^3 + e_2.
\end{equation}
\end{itemize}
\end{PSresult}

In the USR phase, the zero‑mode grows as $\tau^{-3}$, reflecting the rapid amplification of long‑wavelength modes. This growth is responsible for the large tadpole contribution, which eventually scales as $e^{6\Delta\mathcal{N}}$.

\subsection{Matching conditions and integration constants}

We impose continuity of $\mathcal{Q}_{k=0}$ and its derivative at the SRI–USR and USR–SRII transitions. Additionally, we enforce the initial condition at the start of inflation (Bunch–Davies vacuum) and the final condition at the end of inflation. Following the matching procedure described in the main text, the integration constants are fixed to

\begin{PSresult}[title={Integration constants}]
\begin{eqnarray}
c_1 &=& \frac{3iH}{2M_P\sqrt{\epsilon_i}} \big[ -i + (1+i)e^{i} \big] k_{\mathrm{CMB}}^{3/2}, \\[4pt]
c_2 &=& \frac{H}{2M_P\sqrt{ k_{\mathrm{CMB}}^3 \epsilon_i}}, \\[4pt]
d_1 &=& -\frac{3H}{2M_P\sqrt{\epsilon_i}} \big[ 1 + (1-i)e^{i} \big]  k_{\mathrm{CMB}}^{3/2} \, t_i^6, \\[4pt]
d_2 &=& \frac{H}{2M_P\sqrt{ k_{\mathrm{CMB}}^3 \epsilon_i}}, \\[4pt]
e_1 &=& \frac{3iH}{2M_P\sqrt{\epsilon_i}} \big[ -i + (1+i)e^{i} \big] k_{\mathrm{CMB}}^{3/2} \, \frac{t_i^6}{t_e^6}, \\[4pt]
e_2 &=& \frac{H}{2M_P\sqrt{ k_{\mathrm{CMB}}^3 \epsilon_i}}.
\end{eqnarray}
\end{PSresult}

Here $k_{\mathrm{CMB}}$ denotes the super‑horizon reference mode (the CMB pivot scale), $\tau_i$ and $\tau_e$ are the conformal times at the start and end of the USR phase, respectively, and $\epsilon_i$ is the first slow‑roll parameter at the beginning of inflation.

These constants are used in the evaluation of the tadpole kernel $\mathcal{S}(\tau_1,\tau_2;q)$ and ultimately lead to the results presented in Eqs.~\eqref{Two-Point-TadpoleBulk-Result} and \eqref{cubic-tadpole-cmb}.


\end{document}